\documentclass{article}
\usepackage[utf8]{inputenc}
\usepackage{geometry}[left=0.5in,right=0.5in,bottom=0.5in,top=0.5in]
\usepackage[table]{xcolor}
\usepackage{graphicx}
\usepackage{multicol}
\usepackage{multirow}
\usepackage{hyperref}
\hypersetup{colorlinks=true,linkcolor=blue,urlcolor=blue}
\title{Connecticut Redistricting Analysis}
\author{Kyle Evans\footnote{Trinity College, Department of Mathematics}\hspace{8pt}and Katherine T. Chang\footnote{University of Washington, College of Education}}
\date{May 2022}

\begin{document}
\maketitle\vspace{-30pt}
\tableofcontents
\section*{Introduction}
\addcontentsline{toc}{section}{Introduction}
Connecticut passed their new state House of Representatives district plan on November 18, 2021 and passed their new state Senate district plan on November 23, 2021. Each passed unanimously in their 9-person bipartisan Reapportionment Commission; however, the process has been criticized for \href{https://ctnewsjunkie.com/2021/11/29/op-ed-district-maps-there-must-be-a-better-way/}{legislators controlling the process} and for \href{https://www.ctpost.com/opinion/article/Opinion-Redistricting-by-politicians-harmful-to-16691192.php}{the negotiations that serve to protect incumbents}.\vspace{-8pt} \\ \\We were asked by the League of Women Voters of Connecticut to provide descriptive and statistical analyses of the new State House and State Senate maps, with a particular focus on incumbent protection. Thus, the purpose of this analysis is to investigate the extent of incumbent protection in the new Assembly maps while also providing summary data on the new districts. The impact of new districts on incumbents will be analyzed through the location of district borders, with an ensemble analysis to determine if the protection of incumbents constitutes a statistical outlier, and by investigating changes to competitive districts.
\section*{Background}
\addcontentsline{toc}{section}{Background}
\subsection*{Redistricting in Connecticut}\addcontentsline{toc}{subsection}{Redistricting in Connecticut}
When creating new political districts every 10 years, Connecticut's state Constitution requires that districts must be contiguous, representatives live within the district, and ``consistent with federal standards" which includes roughly equal population and in accordance with the Voting Rights Act. In addition, town borders must be maintained as much as possible and only divided to meet other requirements.\vspace{-8pt}\\ \\
In the 2020 Census, Connecticut's population grew by just under 1 percent to 3,605,944 people. On the town level, 68 of the 169 towns saw an increase in population, most notably Stamford, whose population grew by over 12,000 people (over a 10\% increase). The most significant growth in population occurred in Fairfield County, with smaller population increases in the towns surrounding Hartford. While the remaining 101 towns saw a decline in population, no town lost more than 3,800 people over the last decade. Detailed changes in total population by town as well as changes in population by demographics can be explored on CTData's \href{https://www.ctdata.org/census-2020-explorer-page}{interactive map and table}.\vspace{-8pt}\\ \\
In addition, Connecticut passed a law in May 2021 ending the practice of prison gerrymandering, meaning that Census counts were adjusted to count people that were incarcerated as residents of their hometown rather than the town of the prison facility. These changes were made for the purpose of redistricting\footnote{This only applies to Assembly (state House and Senate) maps and not for CT's Congressional map.} and saw Waterbury's population increase by just over 1,000 people while the five towns of Enfield, Suffield, Cheshire, Somers, and Montville all saw their populations decrease by at least 1,000 people. See the state's \href{https://portal.ct.gov/-/media/OPM/CJPPD/CjAbout/SAC-Documents-from-2021-2022/PA21-13_OPM_Summary_Report_20210921.pdf}{OPM report} for additional details about the process and population adjustments for each town. As a result, we have the following ideal populations:\\
\indent State House of Representatives (151 districts): 23,865\\
\indent State Senate (36 districts): 100,099
\subsection*{Ensemble Analysis}\label{MCMC}\addcontentsline{toc}{subsection}{Ensemble Analysis} We can use algorithmic techniques to observe whether the percentage of single incumbents in districts can statistically be considered an outlier within the universe of potential redistricting plans.\vspace{-8pt}\\ \\Current capabilities in statistical analysis and computing facilitate the algorithmic generation of a large number of district map plans (models). Required state redistricting criteria are operationalized as inputs to the statistical models. The output is a set of district plans, the characteristics of which can be observed. One such characteristic is the mean proportion of modeled plans that include a single incumbent in any district. Due to the placement of single incumbents in nearly every district of the actual state House and Senate plans, we are interested in determining the likelihood of that outcome in comparison to the modeled plans generated by an algorithm. \vspace{-8pt}\\ \\
An ensemble analysis employs Markov Chain Monte Carlo (MCMC) methods to consider the range of potential district plans. A redistricting plan can be mathematically modeled as a graph partition, where Census blocks are the vertices and edges represent blocks that border each other and are contained in the same district. Each step of the process involves randomly combining two neighboring districts and then randomly repartitioning them by removing an edge of a spanning tree. This is known as a \textbf{recombination} (ReCom) random walk on the space of graph partitions.\vspace{-8pt}\\ \\
Our ensemble employed the ReCom process across 20,000 steps (where each step represents a district map in our distribution), which previous research (\href{https://mggg.org/VA-report.pdf}{Example 1}, \href{https://hdsr.mitpress.mit.edu/pub/1ds8ptxu/release/5}{Example 2}) has noted as sufficient to reach a steady distribution. Town splits were minimized using an acceptance function that coerced the chain to only accept a next step with fewer town splits than the current step, used alongside a spanning tree algorithm that minimizes town splits, as described \href{https://link.springer.com/article/10.1007/s42001-021-00119-7}{here}.\vspace{-8pt}\\ \\
\href{https://gerrychain.readthedocs.io/en/latest/}{GerryChain} is a Python library that uses MCMC methods to study political redistricting problems by computationally generating redistricting plans from a distribution that accounts for legal rules specific to each state's unique context. A GerryChain user guide can be viewed \href{https://uwescience.github.io/DSSG2021-redistricting-website/guide/}{here}. Our code for each ensemble (House and Senate) is publicly available and linked within the results sections in this report.\vspace{-8pt}\\ \\
Voting Rights Act criteria to ensure the effective representation of linguistic and racial minorities were not included in the ensemble. Election results were also not included, meaning that we are unable to analyze the partisan lean of our modeled districts.
\section*{State House Districts}
\addcontentsline{toc}{section}{State House Districts}
See \hyperref[sec:Appendix 1]{Appendix 1} for complete data on the new State House districts.\footnote{Data from Dave's Redistricting.}\\See \hyperref[sec:Appendix 3]{Appendix 3} for a list of towns and the districts they contain.
\subsection*{Overview}\addcontentsline{toc}{subsection}{Overview}
\textbf{Population Deviation}: Using the 2020 Census data, the largest total population is 24,850 (District 122) and the smallest total population is 22,842 (District 1). This gives a population deviation of 8.41\% which is less than the 10\% generally tolerated by courts.\\ \\
\textbf{Minority Representation}: Using the 2020 Census data, 33.3\% of Connecticut's voting-age population consists of minority populations, including 15\% Hispanic, 11.8\% Black, and 5.3\% Asian. The State House map contains 36 majority-minority districts and an additional 8 districts with minority voting-age populations between 45 and 50 percent.\vspace{-8pt}\\ \\Furthermore, there are 6 districts with a majority Hispanic population: 3 in Hartford, 1 in New Britain, 1 in New Haven, and 1 in Bridgeport. An additional 8 districts have Hispanic populations between 40 and 50 percent. There are 4 districts with a majority Black population: 3 in Hartford and 1 in Bloomfield. An additional 6 districts have Black populations between 40 and 50 percent.\\ \\
\textbf{Partisan Lean}: Using an aggregate of statewide election data during 2016-2020, we can estimate the two-party vote share for each of the new districts. As a result, 86 districts contain greater than 55\% Democratic votes, 11 contain greater than 55\% Republican votes, and 54 are considered competitive with each party having between 45 and 55 percent of the votes. As of the most recent special election in March 2022, Democrats hold 97 seats in the House and Republicans hold the remaining 54 seats.
\subsection*{Incumbency}
\addcontentsline{toc}{subsection}{Incumbency}
The new House map contains incumbents (winners in the November 2020 general election and special elections in the 112th and 145th districts in 2021) in 150 of the 151 districts. The lone exception is in District 42 which moved from the southeastern part of the state to Wilton (and parts of New Canaan and Ridgefield) due to the population shifts in the state. The current incumbent Mike France (R) is running for Congress in Connecticut's 2nd district (eastern half of the state).
\subsubsection*{District Borders}\addcontentsline{toc}{subsubsection}{District Borders}
We examined district borders to assess whether any of the new borders appeared to be tailored in a manner favorable to the incumbent. The very concept and term known as “gerrymandering” derive from a district shape created by then-Governor Eldridge Gerry in 1812 in Massachusetts. The practice of assessing “odd” shapes as one criterion for gerrymandering has persisted.  \vspace{-8pt}\\ \\
We identified several borders that appear to be created simply to ensure that the incumbent still resides within the district. A ``border of note" will be defined as any border that appears tailored to incumbents, generally when a district border, which is not also a town border, is placed very close to an incumbent's residency:\vspace{-8pt}\begin{itemize}
\item District 12 - Geoff Luxenberg (D, Manchester)\vspace{-8pt}
\item District 13 - Jason Doucette (D, Manchester)\vspace{-8pt}
\item District 20 - Kate Farrar (D, West Hartford)\vspace{-8pt}
\item District 25 - Robert Sanchez (D, New Britain)\vspace{-8pt}
\item District 31 - Jill Barry (D, Glastonbury)\vspace{-8pt}
\item District 33 - Brandon Chafee (D, Middletown)\vspace{-8pt}
\item District 46 - Emmett Riley (D, Norwich)\vspace{-8pt}
\item District 47 - Doug Dubitsky (R, Chaplin)\vspace{-8pt}
\item District 82 - Michael Quinn (D, Meriden)\vspace{-8pt}
\item District 88 - Josh Elliott (D, Hamden)\vspace{-8pt}
\item District 100 - Quentin Phipps (D, Middletown)\vspace{-8pt}
\item District 135 - Anne Hughes (D, Easton)\vspace{-8pt}
\item District 139 - Kevin Ryan (D, Montville)\vspace{-8pt}
\item District 142 - Lucy Dathan (D, New Canaan)\vspace{-8pt}
\item District 149 - Kimberly Fiorello (R, Greenwich)
\end{itemize} It is also worth noting the \href{https://ctmirror.org/2022/03/09/complaint-bristol-rep-cara-pavalock-damato-doesnt-have-bona-fide-residence-in-district/}{controversy associated with the borders of District 77 and 78 in Bristol} and the ``address" of the current incumbent.
\subsubsection*{Ensemble Results}
\addcontentsline{toc}{subsubsection}{Ensemble Results}
An ensemble of possible district plans was created to answer the question: Does incumbent count in the 2021 Connecticut State House plan represent an extreme outlier? We are considering incumbency based on the time of redistricting as some incumbents have since announced their retirement from politics or their intention to run for another political position. We also consider incumbent protection to be a district drawn with a single incumbent, as opposed to a district drawn with multiple incumbents or an open seat with no incumbents. \vspace{-8pt}\\ \\
See \hyperref[MCMC]{Ensemble Analysis} in the Background for more information on the methods used. In addition, the model and GerryChain run outputs can be viewed \href{https://github.com/ka-chang/RedistrictingCT/blob/main/03_gerrychain_ensemble.ipynb}{here}.\vspace{-8pt}\\ \\
\includegraphics[width=3in]{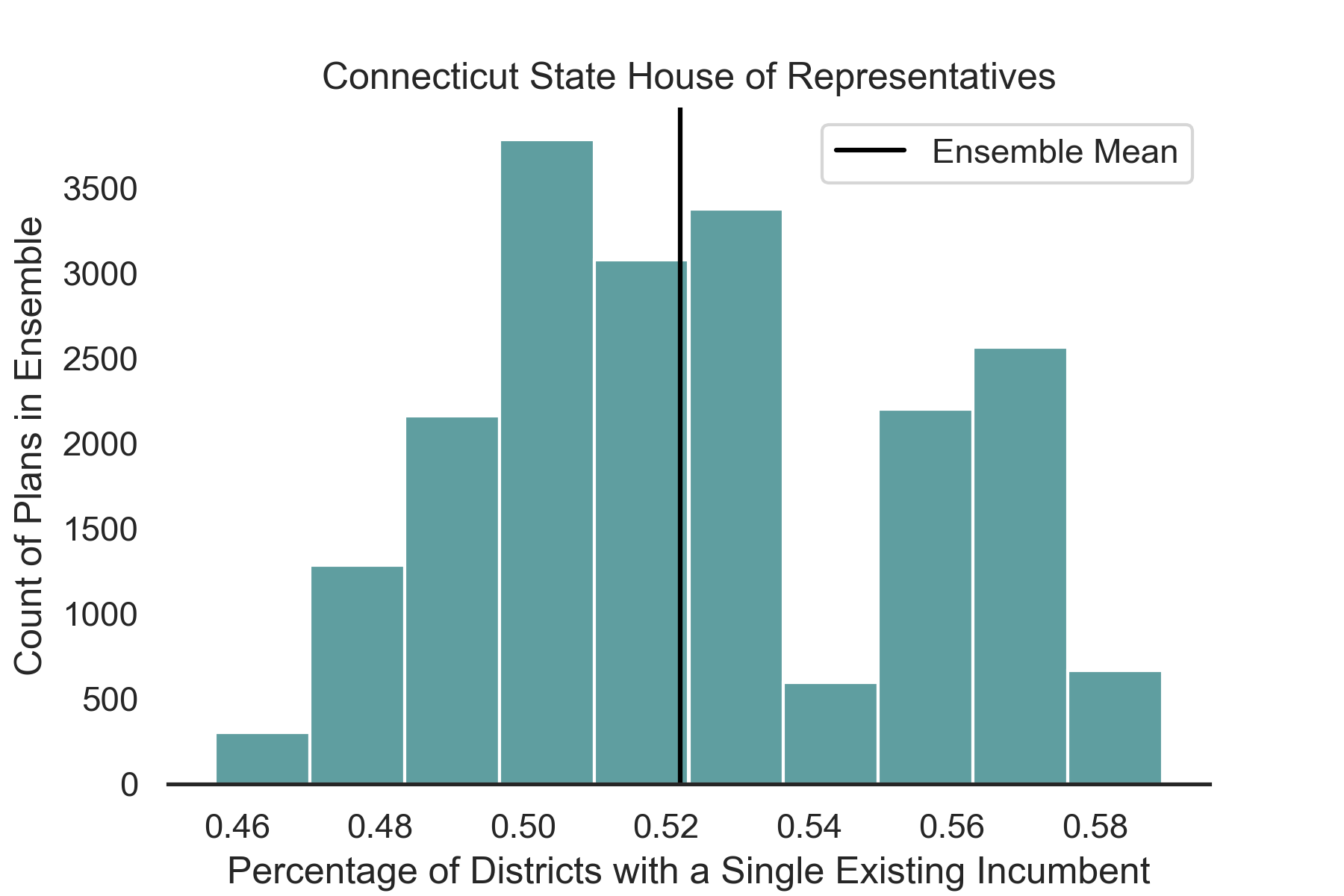}\hspace{10pt}
\includegraphics[width=3in]{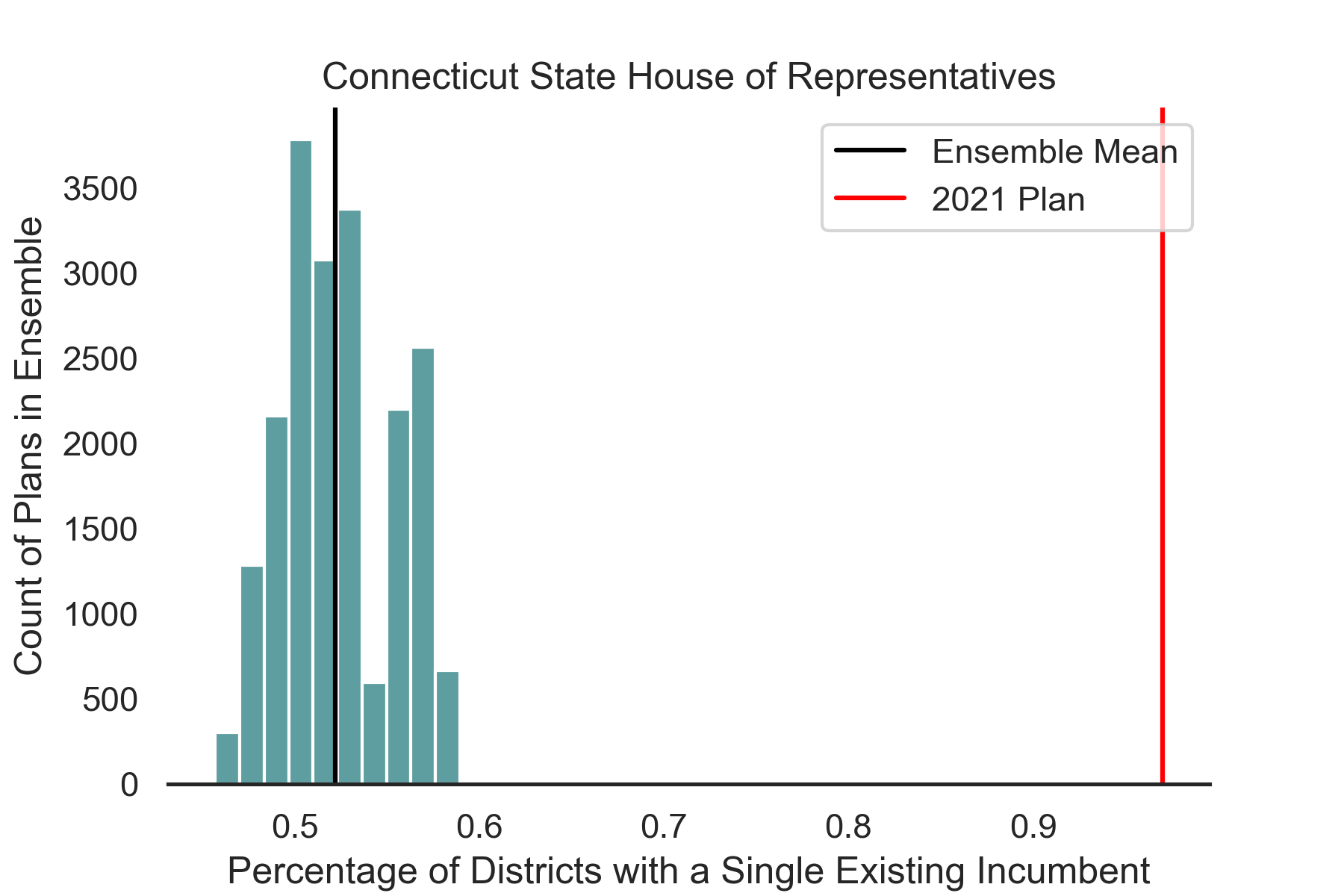}\\ \\
For State House maps, the ensemble (containing 20,000 maps) mean of districts that contain a single incumbent is 52.2\% compared to 97\% in the 2021 State House map. By contrast, on average 47.8\% of the modeled 151 House districts contain either no current incumbent or two or more incumbents within its new boundaries. These results indicate the 2021 CT State House map is an extreme outlier in terms of incumbent placement in newly drawn district boundaries.\\
\begin{center}
\includegraphics[width=3in]{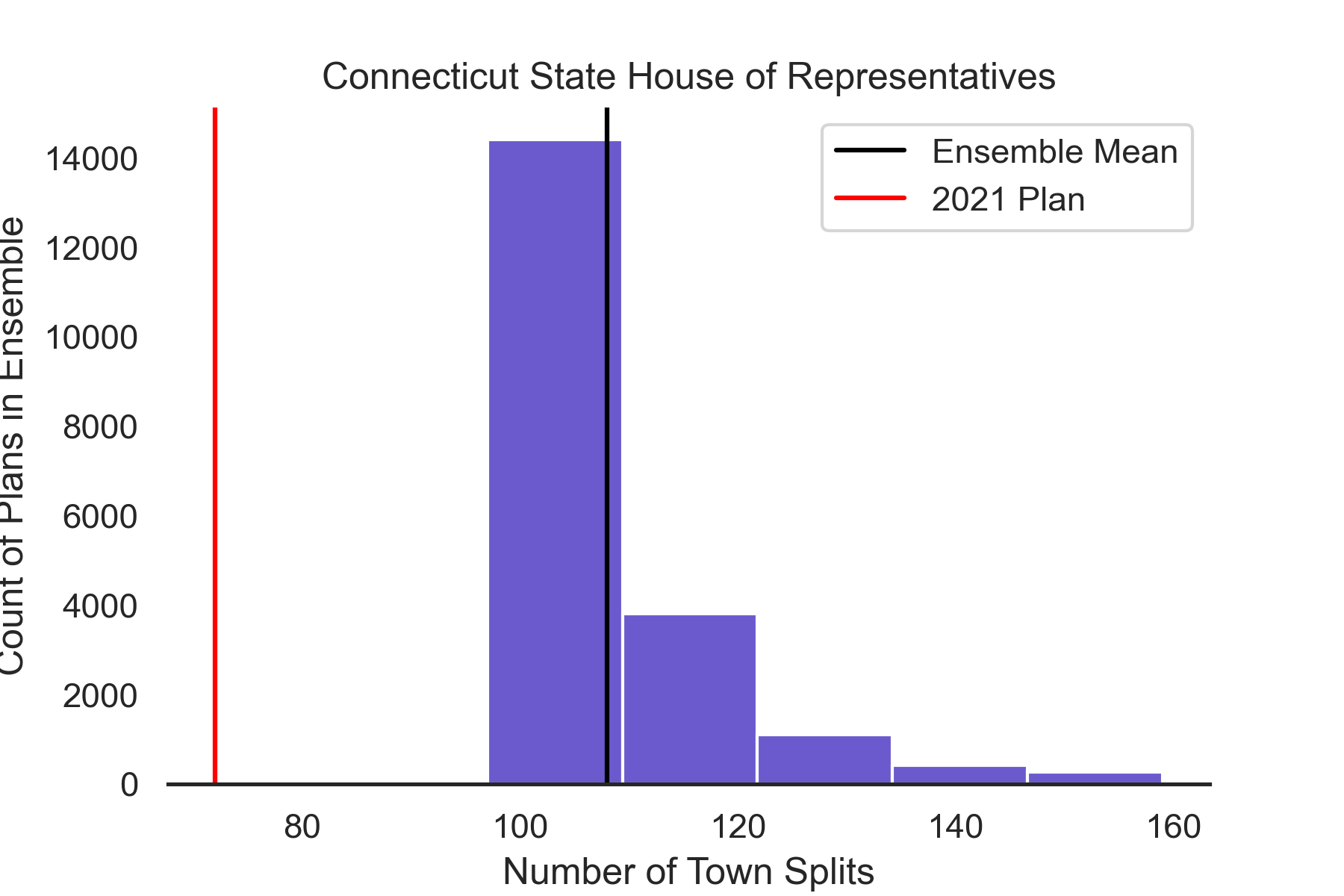}
\end{center}
The ensemble mean of town splits for the potential State House maps is 108 compared to the passed 2021 State House map which has 72 town splits.
\subsubsection*{Changes to Competitive Districts}
\addcontentsline{toc}{subsubsection}{Changes to Competitive Districts}
If incumbents are to be truly ``protected," then they must not only be placed in their own district, but also in a district that they are likely to win. Using the \href{https://ctemspublic.pcctg.net/#/selectTown}{official election results} from the Secretary of State and the new House map, we can analyze the potential impacts on the incumbents from the 10 most competitive elections from November 2020 (margins within 2.5\%).\\\textbf{Note}: State House elections are used to analyze all changes, even if the incumbent was not one of the candidates (applicable to any new additions to districts). All changes in \textbf{bold benefit} the incumbent, changes in \textit{italics disadvantage} the incumbent, and any changes with standard text are neutral.\vspace{-8pt}\\ \\
1) \textbf{Craig Fishbein} - Republican, District 90\vspace{-4pt}\\2020 margin of victory: 7 votes (0.05\%)\\Additions: \textbf{all of Middlefield, part of Wallingford}\vspace{-4pt}\\Subtractions: \textbf{part of Cheshire}\\In addition, Fishbein's 2020 opponent Jim Jinks lives in Cheshire and no longer lives in District 90.\vspace{-8pt}\\ \\
2) \textbf{Liz Linehan} - Democrat, District 103\vspace{-4pt}\\2020 margin of victory: 120 votes (0.95\%)\\Additions: \textbf{part of Hamden}, part of Cheshire\vspace{-4pt}\\Subtractions: \textbf{part of Southington}\vspace{-8pt}\\ \\
3) \textbf{Kathy Kennedy} - Republican, District 119\vspace{-4pt}\\2020 margin of victory: 153 votes (1.1\%)\\Additions: \textbf{part of Orange}\vspace{-4pt}\\Subtractions: None\vspace{-8pt}\\ \\
4) \textbf{Holly Cheeseman} - Republican, District 37\vspace{-4pt}\\2020 margin of victory: 200 votes (1.4\%)\\Additions: \textbf{part of Montville}\vspace{-4pt}\\Subtractions: \textit{part of Salem}\vspace{-8pt}\\ \\
5) \textbf{David Rutigliano} - Republican, District 123\vspace{-4pt}\\2020 margin of victory: 235 votes (1.7\%)\\Additions: \textbf{part of Trumbull}\vspace{-4pt}\\Subtractions: \textbf{part of Trumbull}\\All changes are beneficial to the incumbent due to the election results from the impacted precincts.\vspace{-8pt}\\ \\
6) \textbf{Christine Goupil} - Democrat, District 35\vspace{-4pt}\\2020 margin of victory: 274 votes (1.9\%)\\District 35 remains exactly the same.\vspace{-8pt}\\ \\
7) \textbf{Jennifer Leeper} - Democrat, District 132\vspace{-4pt}\\2020 margin of victory: 293 votes (2.0\%)\\Additions: \textbf{part of Fairfield}\vspace{-4pt}\\Subtractions: \textit{part of Fairfield}\vspace{-8pt}\\ \\
8) \textbf{Greg Howard} - Republican, District 43\vspace{-4pt}\\2020 margin of victory: 295 votes (2.0\%)\\Additions: \textbf{part of Ledyard}\vspace{-4pt}\\Subtractions: \textbf{part of Stonington}\vspace{-8pt}\\ \\
9) \textbf{Kathleen McCarty} - Republican, District 38\vspace{-4pt}\\2020 margin of victory: 345 votes (2.4\%)\\Additions: part of Montville\vspace{-4pt}\\Subtractions: \textbf{part of Montville}\vspace{-8pt}\\ \\
10) \textbf{Robin Green} - Republican, District 55\vspace{-4pt}\\2020 margin of victory: 366 votes (2.4\%)\\Additions: \textit{part of Glastonbury}\vspace{-4pt}\\Subtractions: \textit{part of Bolton}\vspace{-8pt}\\ \\
This analysis extended to the 30 most competitive districts from 2020 shows that 16 of the incumbents benefit from their new districts, 13 see minimal or neutral changes, and only 1 is in a district that is more difficult to win (Robin Green in District 55).
\section*{State Senate Districts}
\addcontentsline{toc}{section}{State Senate Districts}
See \hyperref[sec:Appendix 2]{Appendix 2} for complete data on the new State Senate districts.\footnote{Data from Dave's Redistricting.}\\See \hyperref[sec:Appendix 3]{Appendix 3} for a list of towns and the districts they contain.
\subsection*{Overview}\addcontentsline{toc}{subsection}{Overview}
\textbf{Population Deviation}: Using the 2020 Census data, the largest total population is 105,093 (District 27) and the smallest total population is 95,096 (District 14). This gives a population deviation of 9.99\% which is just within the 10\% generally tolerated by courts.\\ \\
\textbf{Minority Representation}: Using the 2020 Census data, 33.3\% of Connecticut's voting-age population consists of minority populations, including 15\% Hispanic, 11.8\% Black, and 5.3\% Asian. The State Senate map contains 7 majority-minority districts and an additional 5 districts with minority voting-age populations between 45 and 50 percent.\vspace{-8pt}\\ \\ Furthermore, there are 2 districts with Hispanic populations between 45 and 50 percent: 1 in Hartford and 1 in Bridgeport. There is 1 district with a majority Black population in Hartford/Bloomfield and an additional 2 districts with Black populations between 40 and 50 percent: 1 in Bridgeport and 1 in New Haven.\\ \\
\textbf{Partisan Lean}: Using an aggregate of statewide election data during 2016-2020, we can estimate the two-party vote share for each of the new districts. As a result, 23 districts contain greater than 55\% Democratic votes, 1 contains greater than 55\% Republican votes, and 12 are considered competitive with each party having between 45 and 55 percent of the votes. As of the most recent special election in August 2021, Democrats hold 24 seats in the Senate and Republicans hold the remaining 12 seats.
\subsection*{Incumbency}
\addcontentsline{toc}{subsection}{Incumbency}
The new Senate map contains incumbents (winners in the November 2020 general election and special elections in the 27th and 36th districts in 2021) in all 36 of the districts.\\\textbf{Note}: Since the new map passed in November 2021, some incumbents have announced they will not be seeking re-election due to stepping away from politics or running for a different political position.
\subsubsection*{District Borders}\addcontentsline{toc}{subsubsection}{District Borders}
We examined district borders to assess whether any of the new borders appeared to be tailored in a manner favorable to the incumbent. A ``border of note" is again defined as any border that appears tailored to incumbents, generally when a district border, which is not also a town border, is placed very close to an incumbent's residency:\vspace{-8pt}\begin{itemize}
\item District 9 - Matthew Lesser (D, Middletown)\vspace{-8pt}
\item District 22 - Marilyn Moore (D, Bridgeport)
\end{itemize}
\subsubsection*{Ensemble Results}
\addcontentsline{toc}{subsubsection}{Ensemble Results}
An ensemble of possible district plans was created to answer the question: Does incumbent count in the 2021 Connecticut State Senate plan represent an extreme outlier? We are considering incumbency based on the time of redistricting as some incumbents have since announced their retirement from politics or their intention to run for another political position. We also consider incumbent protection to be a district drawn with a single incumbent, as opposed to a district drawn with multiple incumbents or an open seat with no incumbents.\vspace{-8pt}\\ \\ See \hyperref[MCMC]{Ensemble Analysis} in the Background for more information on the methods used. In addition, the model and GerryChain run outputs can be viewed \href{https://github.com/ka-chang/RedistrictingCT/blob/main/03_gerrychain_ensemble.ipynb}{here}.\vspace{-8pt}\\ \\
\includegraphics[width=3in]{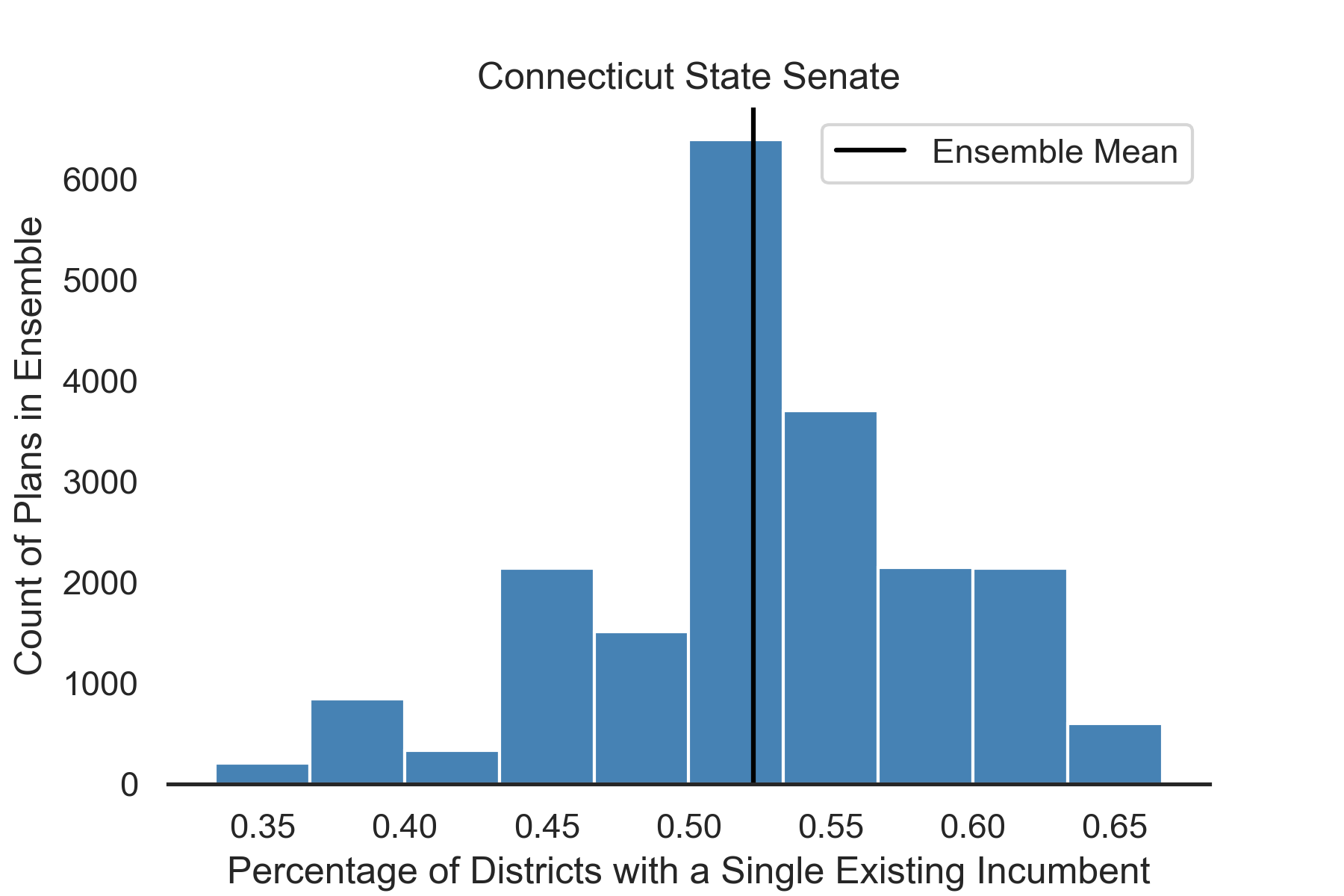}\hspace{10pt}
\includegraphics[width=3in]{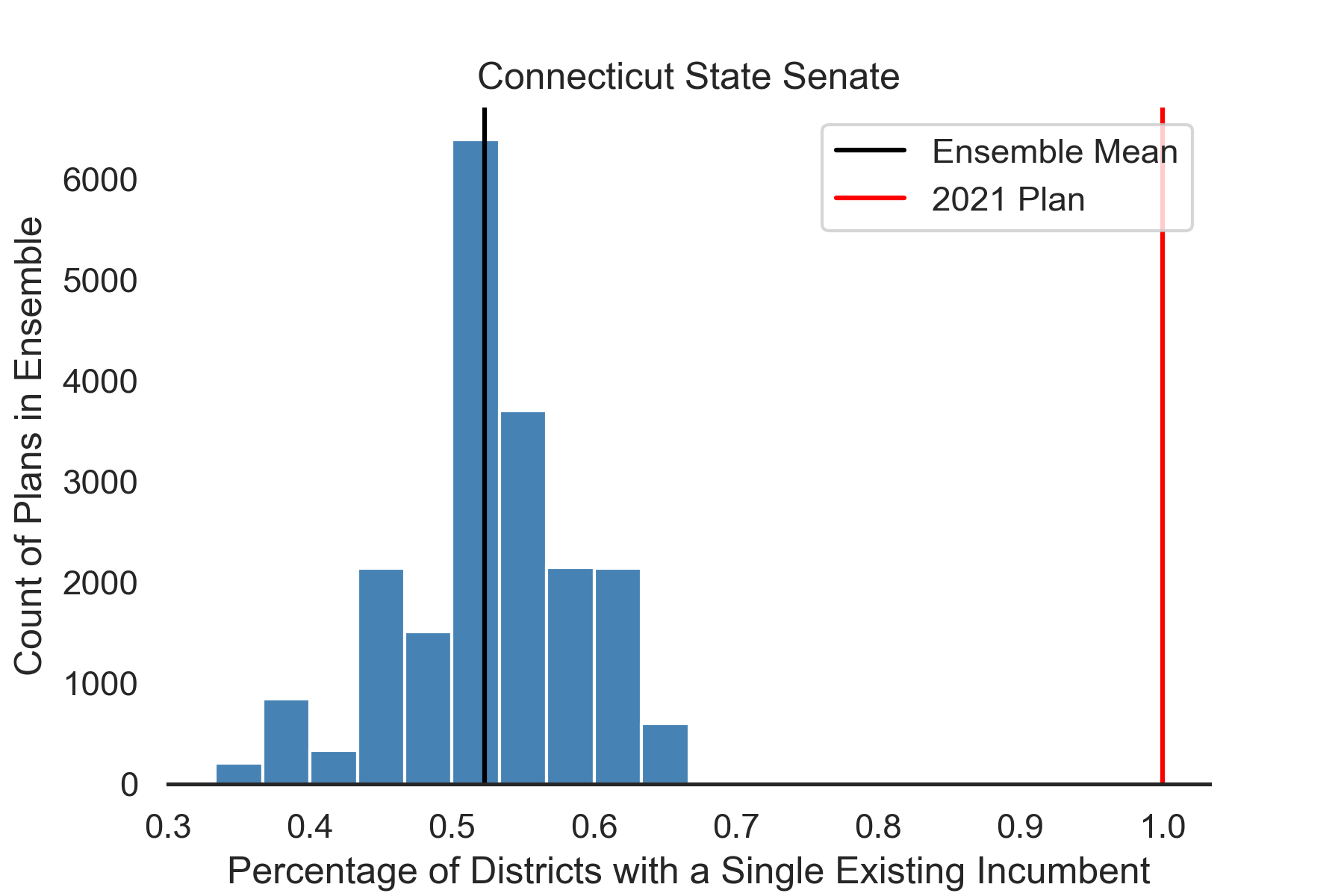}\\ \\
For State Senate maps, the ensemble (containing 20,000 maps) mean of districts that contain a single incumbent is 52.3\% compared to 100\% in the 2021 State Senate map. By contrast, on average 47.7\% of the modeled 36 Senate districts contain either no current incumbent or two or more incumbents within its new boundaries. These results indicate the 2021 CT State Senate map is an extreme outlier in terms of incumbent placement in newly drawn district boundaries.
\begin{center}
\includegraphics[width=3in]{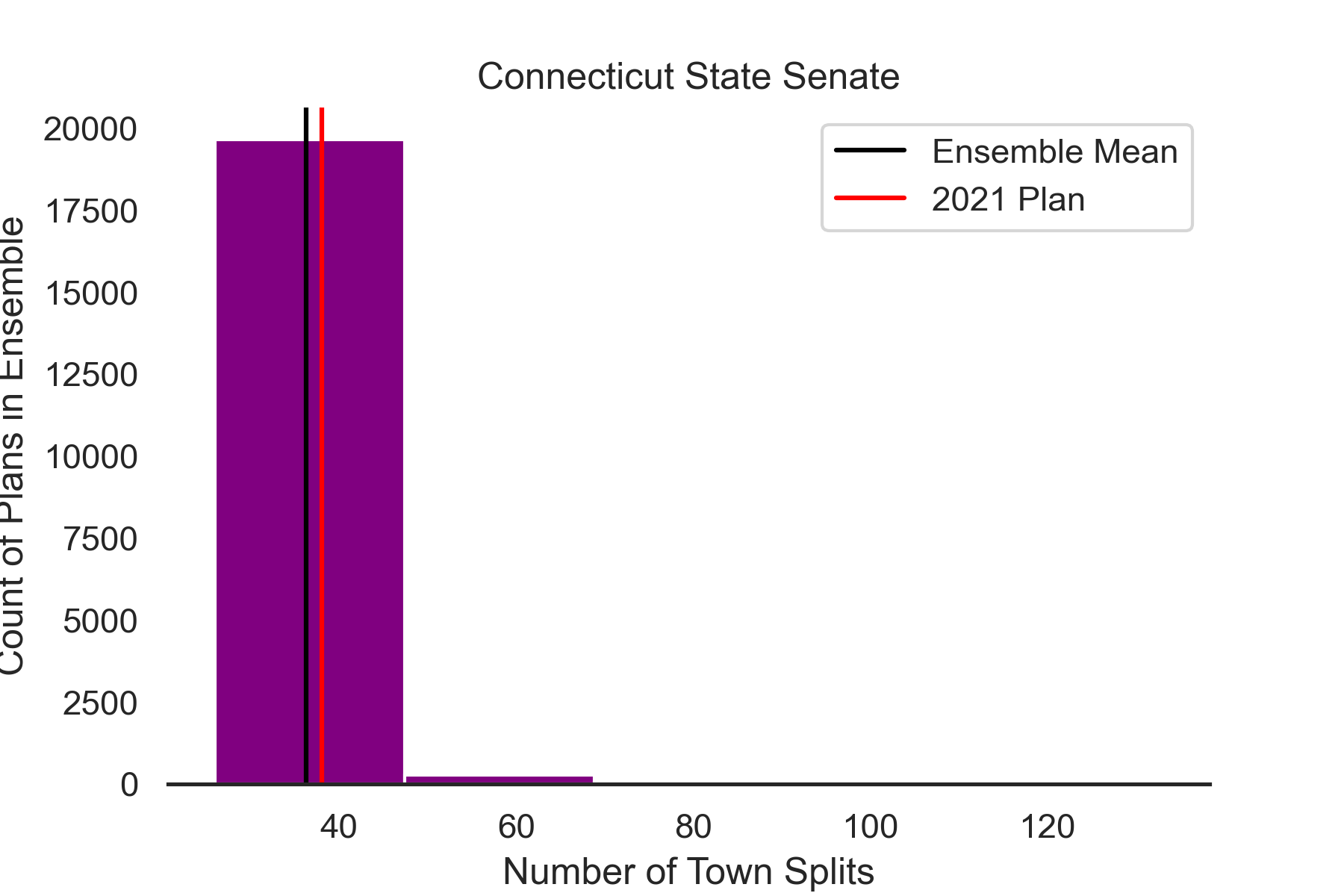}
\end{center}
The ensemble mean of town splits for the potential State Senate maps is 36 compared to the passed 2021 State Senate map which has 38 town splits. These results indicate that the ensemble plans can decrease single incumbent placement and minimize town splits, and that computationally generated redistricting plans can match human-generated plans for town splits in this context.
\subsubsection*{Changes to Competitive Districts}
\addcontentsline{toc}{subsubsection}{Changes to Competitive Districts}
Again, if incumbents are to be truly ``protected," then they must not only be placed in their own district, but also in a district that they are likely to win. Using the \href{https://ctemspublic.pcctg.net/#/selectTown}{official election results} from the Secretary of State and the new Senate map, we can explore the potential impacts on the incumbents from the 9 most competitive elections from November 2020 (margin within 6.0\%).\\\textbf{Note}: State Senate elections are used to analyze all changes, even if the incumbent was not one of the candidates (applicable for any new additions to districts). All changes in \textbf{bold benefit} the incumbent, changes in \textit{italics disadvantage} the incumbent, and any changes with standard text are neutral.\vspace{-8pt}\\ \\
1) \textbf{Dan Champagne} - Republican, District 35\vspace{-4pt}\\2020 margin of victory: 625 votes (1.1\%)\\Additions: \textbf{part of Thompson}\vspace{-4pt}\\Subtractions: \textbf{all of Pomfret}\vspace{-8pt}\\ \\
2) \textbf{Paul Formica} - Republican, District 20\vspace{-4pt}\\2020 margin of victory: 1,177 votes (2.3\%)\\District 20 remains exactly the same.\\Note: Senator Formica was a member of the Reapportionment Commission and also announced\vspace{-4pt}\\his retirement from politics in January 2022.\vspace{-8pt}\\ \\
3) \textbf{Ryan Fazio} - Republican, District 36\vspace{-4pt}\\2020 margin of victory: -1,562 votes (-2.8\%)\vspace{-4pt}\\Note: Senator Fazio won by 2.6\% in a special election in August 2021.\\Additions: \textbf{part of New Canaan}\vspace{-4pt}\\Subtractions: \textbf{part of Stamford}\vspace{-8pt}\\ \\
4) \textbf{Tony Hwang} - Republican, District 28\vspace{-4pt}\\2020 margin of victory: 2,098 votes (3.4\%)\\Additions: part of Bethel\vspace{-4pt}\\Subtractions: \textbf{part of Weston, part of Westport}\vspace{-8pt}\\ \\
5) \textbf{Jorge Cabrera} - Democrat, District 17\vspace{-4pt}\\2020 margin of victory: 2,076 votes (4.3\%)\\Additions: \textbf{part of Hamden}\vspace{-4pt}\\Subtractions: None\vspace{-8pt}\\ \\
6) \textbf{Mae Flexer} - Democrat, District 29\vspace{-4pt}\\2020 margin of victory: 1,878 votes (4.7\%)\\Additions: \textbf{all of Pomfret}\vspace{-4pt}\\Subtractions: \textbf{part of Thompson}\\Note: The only changes in Districts 29 and 35 were with each other and increased their \vspace{-4pt}\\population deviation by 6,000.\vspace{-8pt}\\ \\
7) \textbf{Heather Somers} - Republican, District 18\vspace{-4pt}\\2020 margin of victory: 2,435 votes (4.8\%)\\District 18 remains exactly the same.\vspace{-8pt}\\ \\
8) \textbf{Kevin Witkos} - Republican, District 8\vspace{-4pt}\\2020 margin of victory: 2,883 votes (5.0\%)\\Additions: \textbf{part of Harwinton}\vspace{-4pt}\\Subtractions: None\\Note: Senator Witkos announced his retirement from politics in January 2022.\vspace{-8pt}\\ \\
9) \textbf{Mary Daugherty Abrams} - Democrat, District 13\vspace{-4pt}\\2020 margin of victory: 2,602 votes (5.5\%)\\Additions: \textbf{part of Middletown}\vspace{-4pt}\\Subtractions: \textbf{part of Middlefield}\vspace{-8pt}\\ \\These changes show that in the 9 most competitive districts from November 2020, 7 of the changes benefit the incumbent while the other 2 districts remain exactly the same.
\section*{Conclusion}\addcontentsline{toc}{section}{Conclusion}
While there is no mention in Connecticut's Constitution of the role/use of incumbents in the drawing of Assembly district lines, it is clear that in practice incumbent protection is the priority of the negotiations. This is shown by the strategic placement of district borders close to incumbent addresses, the extreme outliers that the current maps represent with respect to districts with a single incumbent, and many of the most competitive districts being changed to benefit the incumbent (regardless of party). Incumbents are not only being placed in their ``own" district, but are also being drawn into districts that are generally more favorable to their chances of reelection.\vspace{-8pt}\\ \\
Connecticut legislators often champion the state's bipartisan redistricting process and understandably so given that the system works in their favor to retain their elected positions. However, their redistricting process that involves negotiations among the legislators themselves in choosing their own voters, especially at a cost of competitiveness and compactness, ultimately hurts the voters of Connecticut.
\section*{Funding} This work was supported by a grant from the League of Women Voters and their People Powered Fair Maps program.
\section*{References}\addcontentsline{toc}{section}{References}
Clelland, J., Colgate, H., DeFord, D., Malmskog, B., \& Sancier-Barbosa, F. (2022). Colorado \par in context: Congressional redistricting and competing fairness criteria in Colorado. \textit{Journal}\par \textit{of Computational Social Science} \textbf{5}, 189–226. \href{https://doi.org/10.1007/s42001-021-00119-7}{https://doi.org/10.1007/s42001-021-00119-7}\vspace{-8pt}\\ \\
DeFord, D., Duchin, M., \& Solomon, J. (2021). Recombination: A family of Markov chains for \par redistricting. \textit{Harvard Data Science Review.} \href{https://hdsr.mitpress.mit.edu/pub/1ds8ptxu/release/5}{https://doi.org/10.1162/99608f92.eb30390f}\vspace{-8pt}\\ \\
Metric Geometry and Gerrymandering Group. (2018). Comparison of Districting Plans for the \par Virginia House of Delegates, Technical report.  \href{https://mggg.org/VA-report.pdf}{https://mggg.org/VA-report.pdf}
\newpage \newgeometry{left=0.5in,top=0.5in}
\section*{Appendix 1: State House Districts (2022-2032)}\label{sec:Appendix 1}\addcontentsline{toc}{section}{Appendix 1: State House Districts (2022-2032)} VAP = Voting-Age Population, W = White, H = Hispanic, B = Black, A = Asian, PL = Partisan Lean\\ \\
\begin{tabular}{c|c|c|c|c|c|c|p{2.5in}}
    D\# & Pop. & VAP W\% & VAP H\% & VAP B\% & VAP A\% & PL & Towns \\ \hline
    1 & 22,842 & 18.2\% & 22.9\% & \textbf{56.6\%} & 4.7\% & \textcolor{blue}{90\% D} & Hartford \\
    \rowcolor{black!10} 2 & 24,188 & \textbf{72.0\%} & 11.5\% & 5.6\% & 6.9\% & \textcolor{violet}{55\% D} & Danbury, Bethel \\
    3 & 23,031 & 10.2\% & \textbf{60.8\%} & 25.7\% & 4.0\% & \textcolor{blue}{89\% D} & Hartford \\
    \rowcolor{black!10} 4 & 23,748 & 20.2\% & \textbf{50.1\%} & 24.2\% & 8.5\% & \textcolor{blue}{88\% D} & Hartford \\
    5 & 23,667 & 26.5\% & 15.6\% & \textbf{53.4\%} & 5.4\% & \textcolor{blue}{81\% D} & Hartford, Windsor, South Windsor \\
    \rowcolor{black!10} 6 & 23,648 & 18.1\% & \textbf{56.0\%} & 22.8\% & 5.5\% & \textcolor{blue}{85\% D} & Hartford, West Hartford \\
    7 & 22,938 & 12.0\% & 22.0\% & \textbf{66.2\%} & 1.6\% & \textcolor{blue}{95\% D} & Hartford \\
   \rowcolor{black!10} & & & & & & & Coventry, Columbia, Bolton, Tolland, \\ \rowcolor{black!10} \multirow{-2}{*}{8} & \multirow{-2}{*}{23,464} & \multirow{-2}{*}{\textbf{90.8\%}} & \multirow{-2}{*}{3.4\%} & \multirow{-2}{*}{1.4\%} & \multirow{-2}{*}{1.6\%} & \multirow{-2}{*}{\textcolor{violet}{51\% D}} & Lebanon\\ 
    9 & 23,418 & \textbf{51.3\%} & 19.2\% & 20.2\% & 9.2\% & \textcolor{blue}{66\% D} & East Hartford, Manchester \\
    \rowcolor{black!10} 10 & 23,676 & 37.2\% & 30.6\% & 28.6\% & 4.2\% & \textcolor{blue}{72\% D} & East Hartford \\
    11 & 23,512 & 33.4\% & 25.3\% & 28.7\% & 13.5\% & \textcolor{blue}{73\% D} & East Hartford, Manchester \\
    \rowcolor{black!10} 12 & 24,349 & \textbf{58.8\%} & 15.3\% & 17.3\% & 8.4\% & \textcolor{blue}{66\% D} & Manchester \\
    13 & 23,483 & \textbf{67.7\%} & 11.9\% & 12.7\% & 7.1\% & \textcolor{blue}{63\% D} & Manchester, Glastonbury \\
    \rowcolor{black!10} 14 & 23,664 & \textbf{73.2\%} & 4.8\% & 5.0\% & 15.6\% & \textcolor{blue}{60\% D} & South Windsor \\
    15 & 23,775 & 38.7\% & 6.9\% & \textbf{51.3\%} & 3.3\% & \textcolor{blue}{83\% D} & Bloomfield, West Hartford \\
    \rowcolor{black!10} 16 & 24,527 & \textbf{85.8\%} & 4.4\% & 2.8\% & 5.6\% & \textcolor{blue}{58\% D} & Simsbury \\
    17 & 24,278 & \textbf{85.2\%} & 3.7\% & 2.3\% & 7.0\% & \textcolor{violet}{55\% D} & Canton, Avon \\
    \rowcolor{black!10} 18 & 24,398 & \textbf{71.6\%} & 10.1\% & 8.5\% & 8.6\% & \textcolor{blue}{76\% D} & West Hartford \\
    19 & 24,714 & \textbf{82.3\%} & 4.2\% & 3.6\% & 8.5\% & \textcolor{blue}{67\% D} & West Hartford, Avon \\
    \rowcolor{black!10} 20 & 24,801 & \textbf{67.3\%} & 12.5\% & 7.5\% & 11.4\% & \textcolor{blue}{65\% D} & Newington, West Hartford \\
    21 & 23,930 & \textbf{77.0\%} & 4.6\% & 4.0\% & 13.2\% & \textcolor{blue}{56\% D} & Farmington \\
    \rowcolor{black!10} 22 & 23,786 & \textbf{82.1\%} & 7.9\% & 3.8\% & 4.6\% & \textcolor{violet}{51\% R} & Plainville, Farmington, Southington \\
    \multirow{2}{*}{23} & \multirow{2}{*}{23,305} & \multirow{2}{*}{\textbf{90.5\%}} & \multirow{2}{*}{3.9\%} & \multirow{2}{*}{1.1\%} & \multirow{2}{*}{2.4\%} & \multirow{2}{*}{\textcolor{violet}{55\% D}} & Lyme, Old Lyme, Old Saybrook, \\ & & & & & & & Westbrook \\
    \rowcolor{black!10} 24 & 23,798 & 42.2\% & 38.6\% & 18.1\% & 3.0\% & \textcolor{blue}{69\% D} & New Britain \\
    25 & 23,939 & 28.6\% & \textbf{52.7\%} & 19.3\% & 2.3\% & \textcolor{blue}{77\% D} & New Britain \\
    \rowcolor{black!10} 26 & 23,796 & \textbf{56.0\%} & 25.8\% & 13.9\% & 4.6\% & \textcolor{blue}{67\% D} & New Britain \\
    27 & 24,465 & \textbf{71.6\%} & 14.1\% & 7.4\% & 6.3\% & \textcolor{blue}{59\% D} & Newington, New Britain \\
    \rowcolor{black!10} 28 & 24,232 & \textbf{79.6\%} & 11.0\% & 4.6\% & 4.4\% & \textcolor{blue}{56\% D} & Wethersfield \\
    29 & 23,946 & \textbf{72.9\%} & 6.3\% & 4.8\% & 15.1\% & \textcolor{blue}{56\% D} & Rocky Hill, Wethersfield \\
    \rowcolor{black!10} 30 & 23,459 & \textbf{87.7\%} & 4.8\% & 2.0\% & 4.0\% & \textcolor{violet}{53\% R} & Berlin, Southington \\
\end{tabular}
\newpage \newgeometry{left=0.3in,top=0.5in}
\begin{tabular}{c|c|c|c|c|c|c|p{2.5in}}
    D\# & Pop. & VAP W\% & VAP H\% & VAP B\% & VAP A\% & PL & Towns \\ \hline
    31 & 24,104 & \textbf{81.5\%} & 5.5\% & 3.1\% & 8.3\% & \textcolor{blue}{57\% D} & Glastonbury \\
    \rowcolor{black!10} 32 & 23,636 & \textbf{84.2\%} & 5.4\% & 4.5\% & 4.1\% & \textcolor{violet}{53\% D} & Portland, Cromwell \\ 
    33 & 23,887 & \textbf{66.5\%} & 9.6\% & 14.8\% & 8.5\% & \textcolor{blue}{66\% D} & Middletown \\
    \rowcolor{black!10} 34 & 23,186 & \textbf{91.0\%} & 3.2\% & 1.6\% & 1.8\% & \textcolor{violet}{51\% R} & East Hampton, East Haddam, Salem \\
    35 & 23,316 & \textbf{87.6\%} & 6.4\% & 1.8\% & 2.1\% & \textcolor{violet}{52\% D} & Killingworth, Clinton, Westbrook \\
    \rowcolor{black!10} 36 & 23,368 & \textbf{92.0\%} & 3.0\% & 1.4\% & 1.5\% & \textcolor{blue}{56\% D} & Haddam, Chester, Deep River, Essex \\
    37 & 24,671 & \textbf{84.2\%} & 4.7\% & 3.4\% & 6.0\% & \textcolor{blue}{56\% D} & East Lyme, Salem, Montville \\
    \rowcolor{black!10} 38 & 24,691 & \textbf{83.0\%} & 6.5\% & 3.8\% & 4.3\% & \textcolor{violet}{54\% D} & Waterford, Montville \\
    39 & 24,257 & 43.0\% & 32.7\% & 22.4\% & 8.7\% & \textcolor{blue}{78\% D} & New London \\
    \rowcolor{black!10} 40 & 24,477 & \textbf{73.3\%} & 9.6\% & 8.8\% & 6.8\% & \textcolor{blue}{60\% D} & Groton, New London \\
    41 & 24,266 & \textbf{79.9\%} & 7.8\% & 5.6\% & 4.9\% & \textcolor{blue}{63\% D} & Stonington, Groton \\
    \rowcolor{black!10} 42 & 23,519 & \textbf{84.0\%} & 4.5\% & 1.9\% & 8.2\% & \textcolor{blue}{57\% D} & Wilton, New Canaan, Ridgefield \\
    43 & 24,275 & \textbf{86.1\%} & 3.8\% & 2.9\% & 3.0\% & \textcolor{violet}{53\% D} & North Stonington, Stonington, Ledyard \\
    \rowcolor{black!10} 44 & 24,736 & \textbf{87.9\%} & 3.8\% & 2.3\% & 1.9\% & \textcolor{red}{57\% R} & Sterling, Plainfield, Killingly \\
    \multirow{2}{*}{45} & \multirow{2}{*}{24,725} & \multirow{2}{*}{\textbf{86.4\%}} & \multirow{2}{*}{3.9\%} & \multirow{2}{*}{2.7\%} & \multirow{2}{*}{3.0\%} & \multirow{2}{*}{\textcolor{violet}{54\% R}} & Preston, Griswold, Voluntown, Lisbon,\\ & & & & & & & Ledyard \\
    \rowcolor{black!10} 46 & 24,448 & \textbf{52.0\%} & 18.6\% & 17.5\% & 10.3\% & \textcolor{blue}{63\% D} & Norwich \\
    & & & & & & & Chaplin, Scotland, Canterbury, \\
    47 & 24,779 & \textbf{88.4\%} & 4.0\% & 2.4\% & 1.9\% & \textcolor{violet}{55\% R} & Sprague, Brooklyn, Plainfield, Lisbon,\\ & & & & & & & Norwich \\
    \rowcolor{black!10} 48 & 24,437 & \textbf{89.3\%} & 3.9\% & 2.1\% & 1.8\% & \textcolor{violet}{50\%} & Franklin, Bozrah, Colchester, Lebanon \\
    49 & 24,555 & \textbf{53.0\%} & 35.6\% & 6.5\% & 4.8\% & \textcolor{blue}{68\% D} & Windham \\
    \rowcolor{black!10} & & & & & & & Ashford, Eastford, Pomfret, Hampton, \\
    \rowcolor{black!10} \multirow{-2}{*}{50} & \multirow{-2}{*}{24,790} & \multirow{-2}{*}{\textbf{89.7\%}} & \multirow{-2}{*}{3.4\%} & \multirow{-2}{*}{2.2\%} & \multirow{-2}{*}{2.2\%} & \multirow{-2}{*}{\textcolor{violet}{52\% D}} & Mansfield, Woodstock, Brooklyn\\ 
    51 & 24,356 & \textbf{89.9\%} & 3.1\% & 1.9\% & 1.4\% & \textcolor{violet}{54\% R} & Thompson, Putnam, Killingly \\
    \rowcolor{black!10} 52 & 24,736 & \textbf{88.1\%} & 4.2\% & 3.8\% & 1.4\% & \textcolor{red}{56\% R} & Somers, Stafford, Union, Woodstock \\
    53 & 23,437 & \textbf{87.3\%} & 3.8\% & 2.3\% & 4.5\% & \textcolor{violet}{52\% D} & Willington, Tolland, Vernon \\
    \rowcolor{black!10} 54 & 23,875 & \textbf{66.9\%} & 8.4\% & 6.8\% & 17.2\% & \textcolor{blue}{76\% D} & Mansfield \\
    \multirow{2}{*}{55} & \multirow{2}{*}{24,763} & \multirow{2}{*}{\textbf{89.4\%}} & \multirow{2}{*}{3.5\%} & \multirow{2}{*}{1.8\%} & \multirow{2}{*}{3.3\%} & \multirow{2}{*}{\textcolor{violet}{51\% D}} & Andover, Hebron, Marlborough, \\ & & & & & & & Glastonbury, Bolton \\
    \rowcolor{black!10} 56 & 23,166 & \textbf{72.4\%} & 10.1\% & 9.8\% & 6.3\% & \textcolor{blue}{59\% D} & Vernon \\
    57 & 23,721 & \textbf{81.5\%} & 4.6\% & 4.4\% & 7.7\% & \textcolor{violet}{51\% R} & Ellington, East Windsor, Vernon \\
    \rowcolor{black!10} 58 & 23,425 & \textbf{79.1\%} & 8.8\% & 6.9\% & 3.5\% & \textcolor{violet}{54\% D} & Enfield \\ 
    59 & 23,369 & \textbf{78.3\%} & 7.5\% & 8.2\% & 3.9\% & \textcolor{violet}{50\%} & Enfield, East Windsor \\
    \rowcolor{black!10} 60 & 24,270 & \textbf{64.2\%} & 8.1\% & 19.7\% & 7.4\% & \textcolor{blue}{62\% D} & Windsor, Windsor Locks \\
\end{tabular}
\newpage \newgeometry{left=0.3in,top=0.5in}
\begin{tabular}{c|c|c|c|c|c|c|p{2.5in}}
    D\# & Pop. & VAP W\% & VAP H\% & VAP B\% & VAP A\% & PL & Towns \\ \hline
    61 & 23,927 & \textbf{81.1\%} & 6.2\% & 7.2\% & 3.7\% & \textcolor{violet}{51\% R} & Suffield, East Granby, Windsor Locks \\
    \rowcolor{black!10} & & & & & & & Granby, Hartland, Barkhamsted, \\
    \rowcolor{black!10} \multirow{-2}{*}{62} & \multirow{-2}{*}{23,129} & \multirow{-2}{*}{\textbf{92.4\%}} & \multirow{-2}{*}{2.3\%} & \multirow{-2}{*}{1.2\%} & \multirow{-2}{*}{1.5\%} & \multirow{-2}{*}{\textcolor{violet}{54\% R}} & New Hartford\\ 
    63 & 23,720 & \textbf{86.9\%} & 5.5\% & 3.0\% & 2.5\% & \textcolor{red}{59\% R} & Colebrook, Winchester, Torrington \\
    \rowcolor{black!10} & & & & & & & Salisbury, North Canaan, Canaan, \\
    \rowcolor{black!10} 64 & 24,165 & \textbf{90.0\%} & 4.2\% & 1.6\% & 1.9\% & \textcolor{blue}{61\% D} & Norfolk, Sharon, Cornwall, Goshen,\\
    \rowcolor{black!10} & & & & & & &  Kent, Washington \\
    65 & 23,547 & \textbf{74.9\%} & 15.2\% & 5.8\% & 2.6\% & \textcolor{violet}{54\% R} & Torrington \\
    \rowcolor{black!10} & & & & & & & Warren, Morris, Bethlehem, \\
    \rowcolor{black!10} \multirow{-2}{*}{66} & \multirow{-2}{*}{23,613} & \multirow{-2}{*}{\textbf{91.6\%}} & \multirow{-2}{*}{3.1\%} & \multirow{-2}{*}{1.3\%} & \multirow{-2}{*}{1.7\%} & \multirow{-2}{*}{\textcolor{violet}{54\% R}} & Woodbury, Litchfield\\
    67 & 23,454 & \textbf{78.5\%} & 10.2\% & 3.2\% & 3.9\% & \textcolor{violet}{50\%} & New Milford \\
    \rowcolor{black!10} 68 & 23,638 & \textbf{86.8\%} & 5.9\% & 3.5\% & 2.2\% & \textcolor{red}{63\% R} & Watertown, Waterbury \\
    \multirow{2}{*}{69} & \multirow{2}{*}{23,544} & \multirow{2}{*}{\textbf{90.1\%}} & \multirow{2}{*}{3.6\%} & \multirow{2}{*}{1.6\%} & \multirow{2}{*}{3.0\%} & \multirow{2}{*}{\textcolor{violet}{51\% R}} & Roxbury, Bridgewater, New Milford, \\ & & & & & & & Southbury \\
    \rowcolor{black!10} 70 & 23,537 & \textbf{72.5\%} & 12.2\% & 8.7\% & 3.0\% & \textcolor{violet}{55\% R} & Naugatuck \\
    71 & 23,243 & \textbf{68.3\%} & 15.9\% & 10.9\% & 3.4\% & \textcolor{violet}{55\% R} & Middlebury, Waterbury \\
    \rowcolor{black!10} 72 & 23,596 & 22.5\% & 44.2\% & 33.7\% & 1.9\% & \textcolor{blue}{73\% D} & Waterbury \\
    73 & 24,712 & 41.8\% & 31.0\% & 25.3\% & 2.3\% & \textcolor{blue}{58\% D} & Waterbury \\
    \rowcolor{black!10} 74 & 24,027 & 44.2\% & 29.1\% & 23.7\% & 3.6\% & \textcolor{blue}{61\% D} & Waterbury \\
    75 & 24,132 & 23.6\% & 49.7\% & 25.0\% & 2.6\% & \textcolor{blue}{72\% D} & Waterbury \\
    \rowcolor{black!10} & & & & & & & Burlington, Harwinton, Thomaston, \\
    \rowcolor{black!10} \multirow{-2}{*}{76} & \multirow{-2}{*}{23,789} & \multirow{-2}{*}{\textbf{91.4\%}} & \multirow{-2}{*}{3.2\%} & \multirow{-2}{*}{1.2\%} & \multirow{-2}{*}{1.6\%} & \multirow{-2}{*}{\textcolor{red}{60\% R}} & Litchfield\\
    77 & 24,308 & \textbf{80.5\%} & 9.4\% & 5.9\% & 3.2\% & \textcolor{violet}{51\% D} & Bristol \\
    \rowcolor{black!10} 78 & 24,072 & \textbf{83.2\%} & 8.3\% & 4.3\% & 2.3\% & \textcolor{red}{57\% R} & Plymouth, Bristol \\
    79 & 24,336 & \textbf{70.0\%} & 17.4\% & 8.9\% & 2.9\% & \textcolor{violet}{55\% D} & Bristol \\
    \rowcolor{black!10} 80 & 23,222 & \textbf{88.3\%} & 4.7\% & 2.8\% & 2.4\% & \textcolor{red}{64\% R} & Wolcott, Southington \\
    81 & 23,539 & \textbf{88.8\%} & 4.7\% & 2.2\% & 3.0\% & \textcolor{violet}{52\% R} & Southington \\
    \rowcolor{black!10} 82 & 23,735 & \textbf{61.4\%} & 24.5\% & 10.7\% & 3.0\% & \textcolor{blue}{59\% D} & Meriden \\
    83 & 23,365 & \textbf{76.8\%} & 12.6\% & 6.0\% & 3.9\% & \textcolor{violet}{51\% D} & Berlin, Meriden, Cheshire \\
    \rowcolor{black!10} 84 & 23,830 & 35.5\% & 48.2\% & 16.2\% & 1.8\% & \textcolor{blue}{67\% D} & Meriden \\
    85 & 23,420 & \textbf{79.3\%} & 11.9\% & 3.1\% & 4.1\% & \textcolor{violet}{54\% D} & Wallingford \\
    \rowcolor{black!10} 86 & 23,385 & \textbf{88.1\%} & 5.6\% & 2.3\% & 2.6\% & \textcolor{violet}{55\% R} & North Branford, Durham, Guilford \\
    87 & 24,263 & \textbf{83.3\%} & 5.1\% & 4.4\% & 6.2\% & \textcolor{violet}{53\% R} & North Haven \\
    \rowcolor{black!10} 88 & 23,340 & \textbf{72.3\%} & 7.8\% & 12.6\% & 6.6\% & \textcolor{blue}{69\% D} & Hamden \\
    \multirow{2}{*}{89} & \multirow{2}{*}{23,138} & \multirow{2}{*}{\textbf{85.4\%}} & \multirow{2}{*}{5.1\%} & \multirow{2}{*}{3.6\%} & \multirow{2}{*}{4.1\%} & \multirow{2}{*}{\textcolor{red}{56\% R}} & Prospect, Bethany, Waterbury, \\ & & & & & & & Cheshire \\
\end{tabular}
\newpage \newgeometry{left=0.3in,top=0.5in}
\begin{tabular}{c|c|c|c|c|c|c|p{2.5in}}
    D\# & Pop. & VAP W\% & VAP H\% & VAP B\% & VAP A\% & PL & Towns \\ \hline
    \rowcolor{black!10} 90 & 22,966 & \textbf{87.1\%} & 5.2\% & 2.0\% & 4.1\% & \textcolor{violet}{51\% R} & Middlefield, Wallingford \\
    91 & 23,440 & \textbf{50.3\%} & 13.6\% & 29.8\% & 6.6\% & \textcolor{blue}{75\% D} & Hamden \\
    \rowcolor{black!10} 92 & 23,608 & 31.8\% & 19.7\% & 45.3\% & 4.1\% & \textcolor{blue}{90\% D} & New Haven \\
    93 & 23,641 & 26.4\% & 16.6\% & 49.3\% & 8.6\% & \textcolor{blue}{93\% D} & New Haven \\
    \rowcolor{black!10} 94 & 23,923 & 27.6\% & 14.1\% & 46.9\% & 12.0\% & \textcolor{blue}{90\% D} & New Haven, Hamden \\
    95 & 24,398 & 12.2\% & \textbf{59.1\%} & 29.4\% & 2.4\% & \textcolor{blue}{87\% D} & New Haven \\
    \rowcolor{black!10} 96 & 23,979 & \textbf{50.5\%} & 17.5\% & 15.3\% & 16.8\% & \textcolor{blue}{89\% D} & New Haven \\
    97 & 24,142 & 36.1\% & 36.0\% & 25.3\% & 3.4\% & \textcolor{blue}{75\% D} & New Haven \\
    \rowcolor{black!10} 98 & 23,702 & \textbf{89.3\%} & 3.9\% & 1.6\% & 3.7\% & \textcolor{blue}{62\% D} & Guilford, Branford \\
    99 & 23,433 & \textbf{73.9\%} & 15.5\% & 5.7\% & 4.2\% & \textcolor{violet}{51\% R} & East Haven \\
    \rowcolor{black!10} 100 & 23,955 & \textbf{67.9\%} & 10.5\% & 16.4\% & 4.5\% & \textcolor{blue}{65\% D} & Middletown \\
    101 & 22,927 & \textbf{90.6\%} & 3.0\% & 1.2\% & 3.5\% & \textcolor{violet}{54\% D} & Madison, Durham \\
    \rowcolor{black!10} 102 & 23,330 & \textbf{84.0\%} & 5.7\% & 3.4\% & 5.4\% & \textcolor{blue}{59\% D} & Branford \\
    103 & 23,208 & \textbf{80.1\%} & 6.0\% & 7.2\% & 5.4\% & \textcolor{violet}{53\% D} & Cheshire, Hamden, Wallingford \\
    \rowcolor{black!10} 104 & 24,169 & \textbf{61.4\%} & 21.1\% & 14.9\% & 2.4\% & \textcolor{blue}{56\% D} & Ansonia, Derby \\
    105 & 23,338 & \textbf{83.4\%} & 7.9\% & 4.5\% & 2.4\% & \textcolor{red}{57\% R} & Beacon Falls, Seymour, Derby \\
    \rowcolor{black!10} 106 & 24,581 & \textbf{86.4\%} & 5.4\% & 2.9\% & 3.3\% & \textcolor{violet}{53\% D} & Newtown \\
    107 & 24,559 & \textbf{81.3\%} & 7.3\% & 2.7\% & 5.5\% & \textcolor{violet}{51\% R} & Brookfield, Bethel, Newtown \\
    \rowcolor{black!10} & & & & & & & Sherman, New Fairfield, New Milford, \\
    \rowcolor{black!10} \multirow{-2}{*}{108} & \multirow{-2}{*}{23,873} & \multirow{-2}{*}{\textbf{82.1\%}} & \multirow{-2}{*}{8.7\%} & \multirow{-2}{*}{4.2\%} & \multirow{-2}{*}{2.9\%} & \multirow{-2}{*}{\textcolor{violet}{52\% R}} & Danbury \\
    109 & 23,868 & 47.7\% & 29.2\% & 8.9\% & 8.0\% & \textcolor{blue}{59\% D} & Danbury \\
    \rowcolor{black!10} 110 & 23,943 & 31.2\% & 43.1\% & 10.9\% & 5.6\% & \textcolor{blue}{70\% D} & Danbury \\
    111 & 23,248 & \textbf{85.9\%} & 5.1\% & 1.7\% & 5.8\% & \textcolor{blue}{57\% D} & Ridgefield \\
    \rowcolor{black!10} 112 & 24,647 & \textbf{85.1\%} & 6.4\% & 2.8\% & 4.4\% & \textcolor{violet}{53\% R} & Monroe, Easton, Trumbull \\
    113 & 23,204 & \textbf{77.8\%} & 9.6\% & 6.1\% & 5.1\% & \textcolor{red}{56\% R} & Shelton \\
    \rowcolor{black!10} 114 & 23,204 & \textbf{75.1\%} & 7.4\% & 7.3\% & 9.0\% & \textcolor{blue}{56\% D} & Woodbridge, Hamden, Orange, Derby \\
    115 & 23,000 & \textbf{54.6\%} & 21.9\% & 18.7\% & 4.6\% & \textcolor{blue}{63\% D} & West Haven \\
    \rowcolor{black!10} 116 & 22,927 & 34.4\% & 24.3\% & 35.6\% & 6.8\% & \textcolor{blue}{76\% D} & West Haven \\
    117 & 23,835 & \textbf{82.5\%} & 6.7\% & 4.5\% & 5.0\% & \textcolor{violet}{52\% D} & West Haven, Orange, Milford \\
    \rowcolor{black!10} 118 & 23,222 & \textbf{80.3\%} & 7.6\% & 4.4\% & 5.8\% & \textcolor{blue}{56\% D} & Milford \\
    119 & 23,271 & \textbf{82.8\%} & 5.8\% & 3.1\% & 6.9\% & \textcolor{violet}{51\% D} & Milford, Orange \\
    \rowcolor{black!10} 120 & 23,681 & \textbf{69.2\%} & 14.8\% & 12.5\% & 2.9\% & \textcolor{blue}{56\% D} & Stratford \\
\end{tabular}
\newpage \newgeometry{left=0.3in,top=0.5in}
\begin{tabular}{c|c|c|c|c|c|c|p{2.5in}}
    D\# & Pop. & VAP W\% & VAP H\% & VAP B\% & VAP A\% & PL & Towns \\ \hline
    121 & 23,793 & 43.4\% & 24.2\% & 29.1\% & 3.3\% & \textcolor{blue}{69\% D} & Stratford \\
    \rowcolor{black!10} 122 & 24,850 & \textbf{80.2\%} & 7.8\% & 6.0\% & 4.5\% & \textcolor{violet}{54\% R} & Shelton, Stratford, Trumbull \\
    123 & 24,062 & \textbf{81.6\%} & 6.4\% & 3.7\% & 6.6\% & \textcolor{violet}{50\%} & Trumbull \\
    \rowcolor{black!10} 124 & 23,935 & 9.9\% & 42.3\% & 48.2\% & 2.6\% & \textcolor{blue}{89\% D} & Bridgeport \\
    125 & 24,397 & \textbf{84.6\%} & 5.3\% & 2.3\% & 6.7\% & \textcolor{violet}{52\% D} & New Canaan, Stamford, Darien \\
    \rowcolor{black!10} 126 & 24,208 & 14.6\% & 36.4\% & 46.4\% & 2.7\% & \textcolor{blue}{85\% D} & Bridgeport \\
    127 & 23,902 & 35.8\% & 28.4\% & 26.8\% & 3.7\% & \textcolor{blue}{75\% D} & Bridgeport \\
    \rowcolor{black!10} 128 & 24,046 & 6.9\% & \textbf{54.8\%} & 36.8\% & 2.3\% & \textcolor{blue}{89\% D} & Bridgeport \\
    129 & 24,189 & 30.5\% & 34.4\% & 29.9\% & 3.7\% & \textcolor{blue}{78\% D} & Bridgeport \\
    \rowcolor{black!10} 130 & 24,026 & 10.8\% & 44.4\% & 41.9\% & 5.3\% & \textcolor{blue}{88\% D} & Bridgeport \\
    131 & 23,464 & \textbf{83.9\%} & 7.6\% & 4.0\% & 2.1\% & \textcolor{red}{60\% R} & Oxford, Southbury, Naugatuck \\
    \rowcolor{black!10} 132 & 24,784 & \textbf{86.7\%} & 6.2\% & 1.6\% & 4.1\% & \textcolor{blue}{59\% D} & Fairfield \\
    133 & 24,719 & \textbf{68.3\%} & 14.1\% & 8.0\% & 6.3\% & \textcolor{blue}{64\% D} & Fairfield, Bridgeport \\
    \rowcolor{black!10} 134 & 24,138 & \textbf{84.0\%} & 6.6\% & 3.1\% & 5.0\% & \textcolor{violet}{54\% D} & Fairfield, Trumbull \\
    135 & 23,139 & \textbf{86.7\%} & 4.7\% & 1.8\% & 4.9\% & \textcolor{blue}{60\% D} & Redding, Weston, Easton \\
    \rowcolor{black!10} 136 & 23,607 & \textbf{84.3\%} & 5.1\% & 2.0\% & 7.0\% & \textcolor{blue}{68\% D} & Westport \\
    137 & 23,705 & 48.5\% & 29.9\% & 15.1\% & 6.3\% & \textcolor{blue}{70\% D} & Norwalk \\
    \rowcolor{black!10} 138 & 24,579 & \textbf{52.0\%} & 27.3\% & 8.5\% & 7.4\% & \textcolor{blue}{56\% D} & Danbury \\
    139 & 24,677 & \textbf{68.5\%} & 10.6\% & 10.9\% & 7.0\% & \textcolor{violet}{55\% D} & Norwich, Montville, Ledyard \\
    \rowcolor{black!10} 140 & 23,581 & 25.7\% & 45.8\% & 23.7\% & 5.1\% & \textcolor{blue}{79\% D} & Norwalk \\
    141 & 23,473 & \textbf{86.2\%} & 5.1\% & 1.4\% & 5.7\% & \textcolor{violet}{52\% D} & Darien, Norwalk \\
    \rowcolor{black!10} 142 & 23,496 & \textbf{65.2\%} & 17.3\% & 9.9\% & 6.8\% & \textcolor{blue}{61\% D} & Norwalk, New Canaan \\
    143 & 23,218 & \textbf{71.4\%} & 14.3\% & 7.1\% & 6.3\% & \textcolor{blue}{62\% D} & Norwalk, Westport \\
    \rowcolor{black!10} 144 & 24,434 & \textbf{56.3\%} & 16.1\% & 10.3\% & 16.3\% & \textcolor{blue}{67\% D} & Stamford \\
    145 & 23,616 & 23.5\% & 42.8\% & 27.9\% & 6.2\% & \textcolor{blue}{80\% D} & Stamford \\
    \rowcolor{black!10} 146 & 23,652 & \textbf{50.5\%} & 26.7\% & 13.6\% & 9.0\% & \textcolor{blue}{71\% D} & Stamford \\
    147 & 23,895 & \textbf{67.0\%} & 17.1\% & 7.6\% & 7.4\% & \textcolor{blue}{63\% D} & Stamford \\
    \rowcolor{black!10} 148 & 23,593 & 37.7\% & 38.5\% & 16.1\% & 7.6\% & \textcolor{blue}{72\% D} & Stamford \\
    149 & 24,385 & \textbf{79.7\%} & 7.9\% & 3.1\% & 7.3\% & \textcolor{violet}{55\% D} & Greenwich, Stamford \\
    \rowcolor{black!10} 150 & 23,752 & \textbf{68.2\%} & 16.9\% & 4.2\% & 8.1\% & \textcolor{blue}{60\% D} & Greenwich \\
    151 & 23,901 & \textbf{77.1\%} & 9.1\% & 2.2\% & 9.7\% & \textcolor{violet}{55\% D} & Greenwich \\
\end{tabular}
\newpage \newgeometry{left=0.5in,top=0.5in}
\section*{Appendix 2: State Senate Districts (2022-2032)}\label{sec:Appendix 2}\addcontentsline{toc}{section}{Appendix 2: State Senate Districts (2022-2032)} VAP = Voting-Age Population, W = White, H = Hispanic, B = Black, A = Asian, PL = Partisan Lean\\ \\
\begin{tabular}{c|c|c|c|c|c|c|p{2.6in}}
    D\# & Pop. & VAP W\% & VAP H\% & VAP B\% & VAP A\% & PL & Towns \\ \hline
    1 & 95,199 & 31.7\% & 44.6\% & 20.2\% & 5.0\% & \textcolor{blue}{75\% D} & Hartford, Wethersfield \\
    \rowcolor{black!10} 2 & 95,768 & 26.5\% & 15.2\% & \textbf{55.7\%} & 3.8\% & \textcolor{blue}{84\% D} & Windsor, Bloomfield, Hartford \\
    \multirow{2}{*}{3} & \multirow{2}{*}{100,231} & \multirow{2}{*}{\textbf{55.2\%}} & \multirow{2}{*}{17.7\%} & \multirow{2}{*}{18.0\%} & \multirow{2}{*}{8.7\%} & \multirow{2}{*}{\textcolor{blue}{62\% D}} & East Windsor, South Windsor, \\ & & & & & & & East Hartford, Ellington \\
    \rowcolor{black!10} & & & & & & & Glastonbury, Manchester, Bolton, \\
    \rowcolor{black!10} \multirow{-2}{*}{4} & \multirow{-2}{*}{103,085} & \multirow{-2}{*}{\textbf{68.4\%}} & \multirow{-2}{*}{10.4\%} & \multirow{-2}{*}{10.6\%} & \multirow{-2}{*}{9.7\%} & \multirow{-2}{*}{\textcolor{blue}{61\% D}} & Andover \\
    \multirow{2}{*}{5} & \multirow{2}{*}{99,056} & \multirow{2}{*}{\textbf{74.2\%}} & \multirow{2}{*}{8.1\%} & \multirow{2}{*}{6.6\%} & \multirow{2}{*}{9.6\%} & \multirow{2}{*}{\textcolor{blue}{66\% D}} & Burlington, West Hartford, Bloomfield, \\ & & & & & & & Farmington \\
    \rowcolor{black!10} 6 & 98,301 & \textbf{53.6\%} & 30.2\% & 13.3\% & 3.8\% & \textcolor{blue}{62\% D} & New Britain, Berlin, Farmington \\
    & & & & & & & Suffield, Enfield, Somers, East Granby, \\
    7 & 95,996 & \textbf{80.6\%} & 6.8\% & 7.3\% & 3.2\% & \textcolor{violet}{50\%} & Windsor Locks, Granby, Windsor, \\ & & & & & & & Ellington \\
    \rowcolor{black!10} & & & & & & & Norfolk, Colebrook, Hartland, Canton,\\
    \rowcolor{black!10} 8 & 98,681 & \textbf{84.7\%} & 5.8\% & 2.9\% & 4.7\% & \textcolor{violet}{51\% D} & Barkhamsted, New Hartford, Simsbury, \\
    \rowcolor{black!10} & & & & & & & Granby, Harwinton, Torrington\\
    \multirow{2}{*}{9} & \multirow{2}{*}{101,408} & \multirow{2}{*}{\textbf{72.1\%}} & \multirow{2}{*}{8.8\%} & \multirow{2}{*}{9.1\%} & \multirow{2}{*}{9.0\%} & \multirow{2}{*}{\textcolor{blue}{59\% D}} & Newington, Rocky Hill, Cromwell, \\ & & & & & & & Wethersfield, Middletown \\
    \rowcolor{black!10} 10 & 96,926 & 27.8\% & 23.7\% & 43.8\% & 6.0\% & \textcolor{blue}{86\% D} & New Haven, West Haven \\
    11 & 96,099 & 49.4\% & 23.9\% & 17.4\% & 9.6\% & \textcolor{blue}{79\% D} & Hamden, New Haven \\
    \rowcolor{black!10} & & & & & & & Branford, Guilford, Madison,\\
    \rowcolor{black!10} 12 & 97,471 & \textbf{88.3\%} & 4.4\% & 2.1\% & 3.6\% & \textcolor{blue}{56\% D} & Killingworth, Middlefield, Durham, \\
    \rowcolor{black!10} & & & & & & & North Branford, East Haven\\
    \multirow{2}{*}{13} & \multirow{2}{*}{98,961} & \multirow{2}{*}{\textbf{64.5\%}} & \multirow{2}{*}{21.6\%} & \multirow{2}{*}{10.0\%} & \multirow{2}{*}{3.7\%} & \multirow{2}{*}{\textcolor{blue}{58\% D}} & Meriden, Cheshire, Middlefield, \\ & & & & & & & Middletown \\
    \rowcolor{black!10} & & & & & & & Milford, Orange, West Haven, \\
    \rowcolor{black!10} \multirow{-2}{*}{14} & \multirow{-2}{*}{95,096} & \multirow{-2}{*}{\textbf{77.6\%}} & \multirow{-2}{*}{9.0\%} & \multirow{-2}{*}{6.1\%} & \multirow{-2}{*}{5.9\%} & \multirow{-2}{*}{\textcolor{violet}{54\% D}} & Woodbridge \\
    15 & 105,079 & 41.8\% & 33.2\% & 22.6\% & 2.5\% & \textcolor{blue}{58\% D} & Waterbury, Middlebury, Naugatuck \\
    \rowcolor{black!10} & & & & & & & Wolcott, Southington, Prospect, \\
    \rowcolor{black!10} \multirow{-2}{*}{16} & \multirow{-2}{*}{101,256} & \multirow{-2}{*}{\textbf{76.8\%}} & \multirow{-2}{*}{10.7\%} & \multirow{-2}{*}{8.3\%} & \multirow{-2}{*}{3.1\%} & \multirow{-2}{*}{\textcolor{violet}{55\% R}} & Waterbury, Cheshire \\
    \multirow{2}{*}{17} & \multirow{2}{*}{100,684} & \multirow{2}{*}{\textbf{62.9\%}} & \multirow{2}{*}{13.2\%} & \multirow{2}{*}{18.3\%} & \multirow{2}{*}{4.5\%} & \multirow{2}{*}{\textcolor{blue}{57\% D}} & Beacon Falls, Bethany, Ansonia, Derby, \\ & & & & & & & Naugatuck, Hamden, Woodbridge \\
\end{tabular}
\newpage \newgeometry{left=0.3in,top=0.5in}
\begin{tabular}{c|c|c|c|c|c|c|p{2.6in}}
    D\# & Pop. & VAP W\% & VAP H\% & VAP B\% & VAP A\% & PL & Towns \\ \hline
    \rowcolor{black!10} & & & & & & & Plainfield, Sterling, Griswold, Preston, \\
    \rowcolor{black!10} 18 & 99,394 & \textbf{82.9\%} & 5.8\% & 4.3\% & 4.0\% & \textcolor{violet}{53\% D} & Voluntown, North Stonington, \\
    \rowcolor{black!10} & & & & & & & Stonington, Groton\\
    & & & & & & & Marlborough, Columbia, Hebron, \\
    19 & 97,880 & \textbf{82.9\%} & 5.8\% & 4.3\% & 4.0\% & \textcolor{violet}{53\% D} & Lebanon, Franklin, Sprague, Lisbon, \\ & & & & & & & Norwich, Ledyard, Montville \\
    \rowcolor{black!10} & & & & & & & Bozrah, Salem, Old Lyme, East Lyme, \\
    \rowcolor{black!10} 20 & 95,671 & \textbf{74.5\%} & 12.1\% & 8.2\% & 4.2\% & \textcolor{blue}{58\% D} & Waterford, New London, Montville, \\
    \rowcolor{black!10} & & & & & & & Old Saybrook\\
    21 & 103,266 & \textbf{74.7\%} & 11.5\% & 9.0\% & 3.8\% & \textcolor{violet}{51\% R} & Shelton, Monroe, Seymour, Stratford \\
    \rowcolor{black!10} 22 & 100,927 & \textbf{51.9\%} & 21.5\% & 19.4\% & 4.8\% & \textcolor{blue}{62\% D} & Trumbull, Monroe, Bridgeport \\
    23 & 99,780 & 11.7\% & 43.5\% & 42.8\% & 3.1\% & \textcolor{blue}{87\% D} & Bridgeport, Stratford \\
    \rowcolor{black!10} 24 & 103,024 & \textbf{54.3\%} & 25.8\% & 8.1\% & 6.5\% & \textcolor{blue}{57\% D} & Danbury, New Fairfield, Ridgefield \\
    25 & 101,774 & \textbf{55.4\%} & 25.1\% & 13.0\% & 6.0\% & \textcolor{blue}{65\% D} & Norwalk, Darien \\
    \rowcolor{black!10} & & & & & & & Redding, Wilton, Weston, Westport, \\
    \rowcolor{black!10} 26 & 102,354 & \textbf{81.5\%} & 7.4\% & 2.9\% & 6.7\% & \textcolor{blue}{62\% D} & Ridgefield, New Canaan, Darien, \\
    \rowcolor{black!10} & & & & & & & Stamford\\
    27 & 105,093 & 46.7\% & 28.0\% & 15.4\% & 9.6\% & \textcolor{blue}{68\% D} & Darien, Stamford \\
    \rowcolor{black!10} 28 & 104,269 & \textbf{84.1\%} & 6.7\% & 2.7\% & 4.7\% & \textcolor{blue}{56\% D} & Newtown, Easton, Fairfield, Bethel \\
    & & & & & & & Pomfret, Killingly, Putnam, Brooklyn, \\
    29 & 104,418 & \textbf{75.4\%} & 12.0\% & 4.4\% & 6.4\% & \textcolor{blue}{57\% D} & Canterbury, Scotland, Windham, \\ & & & & & & & Mansfield, Thompson \\
    \rowcolor{black!10} & & & & & & & Salisbury, Sharon, Cornwall, Goshen, \\
    \rowcolor{black!10} & & & & & & & North Canaan, Litchfield, Winchester, \\
    \rowcolor{black!10} 30 & 104,615 & \textbf{84.8\%} & 6.8\% & 2.5\% & 3.1\% & \textcolor{violet}{50\%} & Kent, Warren, Sherman, New Milford, \\
    \rowcolor{black!10} & & & & & & & Canaan, Morris, Torrington, Washington, \\
    \rowcolor{black!10} & & & & & & & Bethlehem, New Fairfield, Brookfield \\
    \multirow{2}{*}{31} & \multirow{2}{*}{100,026} & \multirow{2}{*}{\textbf{80.0\%}} & \multirow{2}{*}{10.4\%} & \multirow{2}{*}{5.5\%} & \multirow{2}{*}{2.7\%} & \multirow{2}{*}{\textcolor{violet}{52\% R}} & Thomaston, Plymouth, Bristol, Plainville, \\ & & & & & & & Harwinton \\
    \rowcolor{black!10} & & & & & & & Roxbury, Woodbury, Watertown, Oxford, \\
    \rowcolor{black!10} 32 & 104,966 & \textbf{86.7\%} & 5.5\% & 2.4\% & 3.2\% & \textcolor{red}{56\% R} & Southbury, Seymour, Brookfield, Bethel, \\
    \rowcolor{black!10} & & & & & & & Washington, Bethlehem, Middlebury \\
\end{tabular}
\newpage \newgeometry{left=0.3in,top=0.5in}
\begin{tabular}{c|c|c|c|c|c|c|p{2.6in}}
    D\# & Pop. & VAP W\% & VAP H\% & VAP B\% & VAP A\% & PL & Towns \\ \hline
    \multirow{4}{*}{33} & \multirow{4}{*}{98,384} & \multirow{4}{*}{\textbf{89.6\%}} & \multirow{4}{*}{4.3\%} & \multirow{4}{*}{1.8\%} & \multirow{4}{*}{1.9\%} & \multirow{4}{*}{\textcolor{violet}{53\% D}} & Portland, East Hampton, Colchester, \\ & & & & & & & Haddam, East Haddam, Chester, Lyme, \\ & & & & & & & Deep River, Essex, Clinton, Westbrook, \\ & & & & & & & Old Saybrook \\
    \rowcolor{black!10} & & & & & & & Wallingford, North Haven, Durham, \\
    \rowcolor{black!10} \multirow{-2}{*}{34} & \multirow{-2}{*}{96,609} & \multirow{-2}{*}{\textbf{80.8\%}} & \multirow{-2}{*}{9.5\%} & \multirow{-2}{*}{3.8\%} & \multirow{-2}{*}{4.7\%} & \multirow{-2}{*}{\textcolor{violet}{51\% R}} & North Branford, East Haven \\
    \multirow{4}{*}{35} & \multirow{4}{*}{97,706} & \multirow{4}{*}{\textbf{84.9\%}} & \multirow{4}{*}{4.9\%} & \multirow{4}{*}{3.8\%} & \multirow{4}{*}{4.0\%} & \multirow{4}{*}{\textcolor{violet}{52\% D}} & Stafford, Union, Woodstock, Tolland, \\ & & & & & & & Willington, Ashford, Eastford, Vernon, \\ & & & & & & & Coventry, Chaplin, Hampton, Ellington, \\ & & & & & & & Thompson \\
    \rowcolor{black!10} 36 & 104,113 & \textbf{76.5\%} & 10.3\% & 3.3\% & 8.0\% & \textcolor{blue}{56\% D} & Greenwich, Stamford, New Canaan \\
\end{tabular}
\newpage \newgeometry{left=0.5in,top=0.5in}
\section*{Appendix 3: Districts by Town}\label{sec:Appendix 3}\addcontentsline{toc}{section}{Appendix 3: Districts by Town} C = Congressional District(s), H = State House District(s), S = State Senate District(s)\\Districts that are new to the town are in \textbf{bold}. Districts no longer in the town are in \textit{italics}.\section*{19 Largest Towns (Population over 50,000)}
\begin{tabular}{c|c|c|c|c||c|c}
     Town & Pop. & C 2022 & H 2022 & S 2022 & H 2012 & S 2012 \\ \hline
     Bridgeport & 148,654 & 4 & 124,\hspace{1pt}126,\hspace{1pt}127,\hspace{1pt}128,\hspace{1pt}129,\hspace{1pt}130,\hspace{1pt}\textbf{133} & 22,\hspace{1pt}23 & 124,\hspace{1pt}126,\hspace{1pt}127,\hspace{1pt}128,\hspace{1pt}129,\hspace{1pt}130 & 22,\hspace{1pt}23 \\
     \rowcolor{black!10} Stamford & 135,470 & 4 & \textbf{125},\hspace{1pt}144,\hspace{1pt}145,\hspace{1pt}146,\hspace{1pt}147,\hspace{1pt}148,\hspace{1pt}149 & \textbf{26},\hspace{1pt}27,\hspace{1pt}36 & 144,\hspace{1pt}145,\hspace{1pt}146,\hspace{1pt}147,\hspace{1pt}148,\hspace{1pt}149 & 27,\hspace{1pt}36\\
     New Haven & 134,023 & 3 & 92,\hspace{1pt}93,\hspace{1pt}94,\hspace{1pt}95,\hspace{1pt}96,\hspace{1pt}97 & 10,\hspace{1pt}11 & 92,\hspace{1pt}93,\hspace{1pt}94,\hspace{1pt}95,\hspace{1pt}96,\hspace{1pt}97,\hspace{1pt}\textit{116} & 10,\hspace{1pt}11 \\
     \rowcolor{black!10} Hartford & 121,054 & 1 & 1,\hspace{1pt}3,\hspace{1pt}4,\hspace{1pt}5,\hspace{1pt}6,\hspace{1pt}7 & 1,\hspace{1pt}2 & 1,\hspace{1pt}3,\hspace{1pt}4,\hspace{1pt}5,\hspace{1pt}6,\hspace{1pt}7 & 1,\hspace{1pt}2\\
     Waterbury & 114,403 & 3,\hspace{1pt}5 & \textbf{68},\hspace{1pt}71,\hspace{1pt}72,\hspace{1pt}73,\hspace{1pt}74,\hspace{1pt}75,\hspace{1pt}\textbf{89} & 15,\hspace{1pt}16 & 71,\hspace{1pt}72,\hspace{1pt}73,\hspace{1pt}74,\hspace{1pt}75 & 15,\hspace{1pt}16\\
     \rowcolor{black!10} Norwalk & 91,184 & 4 & 137,\hspace{1pt}140,\hspace{1pt}141,\hspace{1pt}142,\hspace{1pt}143 & 25 & 137,\hspace{1pt}140,\hspace{1pt}141,\hspace{1pt}142,\hspace{1pt}143 & 25\\
     Danbury & 86,518 & 5 & 2,\hspace{1pt}108,\hspace{1pt}109,\hspace{1pt}110,\hspace{1pt}138 & 24 & 2,\hspace{1pt}\textit{107},\hspace{1pt}108,\hspace{1pt}109,\hspace{1pt}110,\hspace{1pt}138 & 24\\
     \rowcolor{black!10} New Britain & 74,135 & 5 & 24,\hspace{1pt}25,\hspace{1pt}26,\hspace{1pt}\textbf{27} & 6 & \textit{22},\hspace{1pt}24,\hspace{1pt}25,\hspace{1pt}26 & 6\\
     West Hartford & 64,083 & 1 & \textbf{6,}\hspace{1pt}\textbf{15},\hspace{1pt}18,\hspace{1pt}19,\hspace{1pt}20 & 5 & 18,\hspace{1pt}19,\hspace{1pt}20 & 5\\
     \rowcolor{black!10} Greenwich & 63,518 & 4 & 149,\hspace{1pt}150,\hspace{1pt}151 & 36 & 149,\hspace{1pt}150,\hspace{1pt}151 & 36\\
     Fairfield & 61,512 & 4 & 132,\hspace{1pt}133,\hspace{1pt}134 & 28 & 132,\hspace{1pt}133,\hspace{1pt}134 & 28 \\
     \rowcolor{black!10}Hamden & 61,169 & 3 & 88,\hspace{1pt}91,\hspace{1pt}94,\hspace{1pt}\textbf{103,}\hspace{1pt}\textbf{114} & 11,\hspace{1pt}17 & 88,\hspace{1pt}91,\hspace{1pt}94 & 11,\hspace{1pt}17 \\
     Meriden & 60,850 & 5 & 82,\hspace{1pt}83,\hspace{1pt}84 & 13 & 82,\hspace{1pt}83,\hspace{1pt}84 & 13 \\
     \rowcolor{black!10} Bristol & 60,833 & 1 & 77,\hspace{1pt}78,\hspace{1pt}79 & 31 & 77,\hspace{1pt}78,\hspace{1pt}79 & 31 \\
     Manchester & 59,713 & 1 & 9,\hspace{1pt}11,\hspace{1pt}12,\hspace{1pt}13 & 4 & 9,\hspace{1pt}11,\hspace{1pt}12,\hspace{1pt}13 & 4 \\
     \rowcolor{black!10}West Haven & 55,584 & 3 & 115,\hspace{1pt}116,\hspace{1pt}117 & 10,\hspace{1pt}14 & 115,\hspace{1pt}116,\hspace{1pt}117 & 10,\hspace{1pt}14 \\
     Stratford & 52,355 & 3 & 120,\hspace{1pt}121,\hspace{1pt}122 & 21,\hspace{1pt}23 & 120,\hspace{1pt}121,\hspace{1pt}122 & 21,\hspace{1pt}23 \\
     \rowcolor{black!10}Milford & 52,044 & 3 & 117,\hspace{1pt}118,\hspace{1pt}119 & 14 & 117,\hspace{1pt}118,\hspace{1pt}119 & 14 \\
     East Hartford & 51,045 & 1 & 9,\hspace{1pt}10,\hspace{1pt}11 & 3 & 9,\hspace{1pt}10,\hspace{1pt}11 & 3 \\
\end{tabular}\\
\newpage \newgeometry{left=0.75in,top=0.75in}\section*{Remaining 150 Towns}
\begin{tabular}{c|c|c|c|c||c|c}
    Town & Pop. & C 2022 & H 2022 & S 2022 & H 2012 & S 2012 \\ \hline
    Andover & 3,151 & 2 & 55 & 4 & 55 & 4 \\
    \rowcolor{black!10} Ansonia & 18,918 & 3 & 104 & 17 & 104 & 17 \\
    Ashford & 4,191 & 2 & \textbf{50} & 35 & \textit{53} & 35 \\
    \rowcolor{black!10} Avon & 18,932 & 5 & 17,\hspace{1pt}19 & 8 & 17,\hspace{1pt}19 & 8 \\
    Barkhamsted & 3,647 & 1 & 62 & 8 & 62 & 8 \\
    \rowcolor{black!10} Beacon Falls & 6,000 & 3 & 105 & 17 & 105 & 17 \\
    Berlin & 20,175 & 1 & 30,\hspace{1pt}83 & 6 & 30,\hspace{1pt}83 & 6 \\
    \rowcolor{black!10} Bethany & 5,297 & 3 & 89 & 17 & 89 & 17 \\
    Bethel & 20,358 & 5 & 2,\hspace{1pt}107 & \textbf{28,\hspace{1pt}32} & 2,\hspace{1pt}107 & \textit{24},\hspace{1pt}\textit{26}\\
    \rowcolor{black!10} Bethlehem & 3,385 & 5 & 66 & \textbf{30},\hspace{1pt}32 & 66 & 32 \\
    Bloomfield & 21,535 & 1 & 15 & 2,\hspace{1pt}5 & 15 & 2,\hspace{1pt}5 \\
    \rowcolor{black!10} Bolton & 4,858 & 2 & \textbf{8},\hspace{1pt}55 & 4 & 55 & 4 \\
    Bozrah & 2,429 & 2 & \textbf{48} & 20 & \textit{139} & 20 \\
    \rowcolor{black!10} Branford & 28,273 & 3 & 98,\hspace{1pt}102 & 12 & 98,\hspace{1pt}102 & 12 \\
    Bridgewater & 1,662 & 5 & 69 & 32 & 69 & 32 \\
    \rowcolor{black!10} Brookfield & 17,528 & 5 & 107 & \textbf{28},\hspace{1pt}30 & 107 & 30 \\
    Brooklyn & 8,450 & 2 & \textbf{47},\hspace{1pt}50 & 29 & 50 & 29 \\
    \rowcolor{black!10}Burlington & 9,519 & 5 & 76 & 5 & 76 & 5 \\
    Canaan & 1,080 & 5 & 64 & 30 & 64 & 30 \\
    \rowcolor{black!10}Canterbury & 5,045 & 2 & 47 & 29 & 47 & 29 \\
    Canton & 10,124 & 5 & 17 & 8 & 17 & 8 \\
    \rowcolor{black!10}Chaplin & 2,151 & 2 & 47 & 35 & 47 & 35 \\
    Cheshire & 28,733 & 5 & \textbf{83},\hspace{1pt}89,\hspace{1pt}103 & 13,\hspace{1pt}16 & 89,\hspace{1pt}\textit{90},\hspace{1pt}103 & 13,\hspace{1pt}16 \\
    \rowcolor{black!10}Chester & 3,749 & 2 & 36 & 33 & 36 & 33 \\
    Clinton & 13,185 & 2 & 35 & 33 & 35 & 33 \\
    \rowcolor{black!10}Colchester & 15,555 & 2 & 48 & 33 & \textit{34},\hspace{1pt}48 & 33 \\
    Colebrook & 1,361 & 1 & 63 & 8 & 63 & 8 \\
    \rowcolor{black!10}Columbia & 5,272 & 2 & 8 & 19 & 8 & 19 \\
    Cornwall & 1,567 & 5 & 64 & 30 & 64 & 30 \\
    \rowcolor{black!10}Coventry & 12,235 & 2 & 8 & 35 & 8 & 35 \\
\end{tabular}
\newpage \newgeometry{left=0.6in,top=0.75in}
\begin{tabular}{c|c|c|c|c||c|c}
    Town & Pop. & C 2022 & H 2022 & S 2022 & H 2012 & S 2012 \\ \hline
    Cromwell & 14,225 & 1 & 32 & 9 & 32 & 9 \\
    \rowcolor{black!10}Darien & 21,499 & 4 & \textbf{125},\hspace{1pt}141 & 25,\hspace{1pt}\textbf{26},\hspace{1pt}27 & 141,\hspace{1pt}\textit{147} & 25,\hspace{1pt}27 \\
    Deep River & 4,415 & 2 & 36 & 33 & 36 & 33 \\
    \rowcolor{black!10}Derby & 12,325 & 3 & 104,\hspace{1pt}105,\hspace{1pt}114 & 17 & 104,\hspace{1pt}105,\hspace{1pt}114 & 17 \\
    Durham & 7,152 & 3 & 86,\hspace{1pt}101 & 12,\hspace{1pt}34 & 86,\hspace{1pt}101 & 12,\hspace{1pt}34 \\
    \rowcolor{black!10}East Granby & 5,214 & 1 & 61 & 7 & 61 & 7 \\
    East Haddam & 8,875 & 2 & 34 & 33 & 34 & 33 \\
    \rowcolor{black!10}East Hampton & 12,717 & 2 & 34 & 33 & 34 & 33 \\
    East Haven & 27,943 & 3 & \textbf{86},\hspace{1pt}99 & \textbf{12},\hspace{1pt}34 & \textit{96},\hspace{1pt}99 & 34 \\
    \rowcolor{black!10}East Lyme & 18,693 & 2 & 37 & 20 & 37 & 20 \\
    East Windsor & 11,190 & 1 & 57,\hspace{1pt}59 & 3 & 57,\hspace{1pt}59 & 3 \\
    \rowcolor{black!10}Eastford & 1,649 & 2 & 50 & 35 & 50 & 35 \\
    Easton & 7,605 & 4 & \textbf{112},\hspace{1pt}135 & 28 & 135 & 28 \\
    \rowcolor{black!10}Ellington & 16,426 & 2 & 57 & 3,\hspace{1pt}\textbf{7},\hspace{1pt}35 & 57 & 3,\hspace{1pt}35 \\
    Enfield & 42,141 & 2 & 58,\hspace{1pt}59 & 7 & 58,\hspace{1pt}59 & 7 \\
    \rowcolor{black!10}Essex & 6,733 & 2 & 36 & 33 & 36 & 33 \\
    Farmington & 26,712 & 5 & 21,\hspace{1pt}\textbf{22} & 5,\hspace{1pt}6 & \textit{19},\hspace{1pt}21 & 5,\hspace{1pt}6 \\
    \rowcolor{black!10}Franklin & 1,863 & 2 & \textbf{48} & 19 & \textit{47} & 19 \\
    Glastonbury & 35,159 & 1,\hspace{1pt}2 & 13,\hspace{1pt}31,\hspace{1pt}\textbf{55} & 4 & 13,\hspace{1pt}31 & 4 \\
    \rowcolor{black!10}Goshen & 3,150 & 5 & 64 & 30 & \textit{63},\hspace{1pt}64 & 30 \\
    Granby & 10,903 & 1 & 62 & 7,\hspace{1pt}8 & 62 & 7,\hspace{1pt}8 \\
    \rowcolor{black!10}Griswold & 11,402 & 2 & 45 & 18 & 45 & 18 \\
    Groton & 38,411 & 2 & 40,\hspace{1pt}41 & 18 & 40,\hspace{1pt}41 & 18 \\
    \rowcolor{black!10}Guilford & 22,073 & 3 & 86,\hspace{1pt}98 & 12 & 86,\hspace{1pt}98 & 12 \\
    Haddam & 8,452 & 2 & 36 & 33 & 36 & 33 \\
    \rowcolor{black!10}Hampton & 1,728 & 2 & \textbf{50} & 35 & \textit{47} & 35 \\
    Hartland & 1,901 & 1 & 62 & 8 & 62 & 8 \\
    \rowcolor{black!10}Harwinton & 5,484 & 5 & 76 & 8,\hspace{1pt}31 & 76 & 8,\hspace{1pt}31 \\
    Hebron & 9,098 & 2 & 55 & 19 & 55 & 19 \\
    \rowcolor{black!10}Kent & 3,019 & 5 & 64 & 30 & 64 & 30 \\
\end{tabular}
\newpage
\begin{tabular}{c|c|c|c|c||c|c}
    Town & Pop. & C 2022 & H 2022 & S 2022 & H 2012 & S 2012 \\ \hline
    Killingly & 17,752 & 2 & 44,\hspace{1pt}51 & 29 & 44,\hspace{1pt}51 & 29 \\
    \rowcolor{black!10}Killingworth & 6,174 & 2 & 35 & 12 & 35 & 12 \\
    Lebanon & 7,142 & 2 & \textbf{8},\hspace{1pt}48 & 19 & \textit{47},\hspace{1pt}48 & 19 \\
    \rowcolor{black!10}Ledyard & 15,413 & 2 & \textbf{43,}\hspace{1pt}\textbf{45,}\hspace{1pt}\textbf{139} & 19 & \textit{40},\hspace{1pt}\textit{42} & 19 \\
    Lisbon & 4,195 & 2 & 45,\hspace{1pt}47 & 19 & 45,\hspace{1pt}47 & 19 \\
    \rowcolor{black!10}Litchfield & 8,192 & 5 & 66,\hspace{1pt}76 & 30 & 66,\hspace{1pt}76 & 30 \\
    Lyme & 2,352 & 2 & 23 & 33 & 23 & 33 \\
    \rowcolor{black!10}Madison & 17,691 & 2 & 101 & 12 & 101 & 12 \\
    Mansfield & 25,892 & 2 & \textbf{50},\hspace{1pt}54 & 29 & \textit{48},\hspace{1pt}54 & 29 \\
    \rowcolor{black!10}Marlborough & 6,133 & 2 & 55 & 19 & 55 & 19 \\
    Middlebury & 7,574 & 5 & 71 & 15,\hspace{1pt}32 & 71 & 15,\hspace{1pt}32 \\
    \rowcolor{black!10}Middlefield & 4,217 & 3 & \textbf{90} & \textbf{12},\hspace{1pt}13 & \textit{82} & 13 \\
    Middletown & 47,717 & 1,\hspace{1pt}3 & 33,\hspace{1pt}100 & 9,\hspace{1pt}13 & 33,\hspace{1pt}100 & 9,\hspace{1pt}13 \\
    \rowcolor{black!10}Monroe & 18,825 & 4 & 112 & 21,\hspace{1pt}22 & 112 & 21,\hspace{1pt}22 \\
    Montville & 18,387 & 2 & \textbf{37},\hspace{1pt}38,\hspace{1pt}139 & 19,\hspace{1pt}20 & 38,\hspace{1pt}\textit{42},\hspace{1pt}139 & 19,\hspace{1pt}20 \\
    \rowcolor{black!10}Morris & 2,256 & 5 & 66 & 30 & 66 & 30 \\
    Naugatuck & 31,519 & 3 & 70,\hspace{1pt}131 & 15,\hspace{1pt}17 & 70,\hspace{1pt}131 & 15,\hspace{1pt}17 \\
    \rowcolor{black!10}New Canaan & 20,622 & 4 & \textbf{42},\hspace{1pt}125,\hspace{1pt}142 & 26,\hspace{1pt}36 & 125,\hspace{1pt}142 & 26,\hspace{1pt}36 \\
    New Fairfield & 13,579 & 5 & 108 & 24,\hspace{1pt}\textbf{30} & 108,\hspace{1pt}\textit{138} & 24 \\
    \rowcolor{black!10}New Hartford & 6,658 & 1 & 62 & 8 & 62 & 8 \\
    New London & 27,367 & 2 & 39,\hspace{1pt}\textbf{40} & 20 & 39,\hspace{1pt}\textit{41} & 20 \\
    \rowcolor{black!10}New Milford & 28,115 & 5 & 67,\hspace{1pt}\textbf{69},\hspace{1pt}108 & 30 & 67,\hspace{1pt}108 & 30 \\
    Newington & 30,536 & 1 & \textbf{20},\hspace{1pt}27 & 9 & \textit{24},\hspace{1pt}27,\hspace{1pt}\textit{29} & 9 \\
    \rowcolor{black!10}Newtown & 27,173 & 5 & 106,\hspace{1pt}\textbf{107} & 28 & \textit{2},\hspace{1pt}106,\hspace{1pt}\textit{112} & 28 \\
    Norfolk & 1,588 & 5 & 64 & \textbf{30} & 64 & \textit{8} \\
    \rowcolor{black!10}North Branford & 13,544 & 3 & 86 & 12,\hspace{1pt}\textbf{34} & 86 & 12 \\
    North Canaan & 3,211 & 5 & 64 & 30 & 64 & 30 \\
    \rowcolor{black!10}North Haven & 24,253 & 3 & 87 & 34 & 87 & \textit{11},\hspace{1pt}34 \\
    North Stonington & 5,149 & 2 & 43 & 18 & 43 & 18 \\
    \rowcolor{black!10}Norwich & 40,125 & 2 & 46,\hspace{1pt}47,\hspace{1pt}139 & 19 & 46,\hspace{1pt}47,\hspace{1pt}139 & 19 \\
\end{tabular}
\newpage
\begin{tabular}{c|c|c|c|c||c|c}
    Town & Pop. & C 2022 & H 2022 & S 2022 & H 2012 & S 2012 \\ \hline
    Old Lyme & 7,628 & 2 & 23 & 20 & 23 & 20 \\
    \rowcolor{black!10}Old Saybrook & 10,481 & 2 & 23 & 20,\hspace{1pt}33 & 23 & 20,\hspace{1pt}33 \\
    Orange & 14,280 & 3 & 114,\hspace{1pt}117,\hspace{1pt}119 & 14 & 114,\hspace{1pt}117,\hspace{1pt}119 & 14 \\
    \rowcolor{black!10}Oxford & 12,706 & 4 & 131 & 32 & 131 & 32 \\
    Plainfield & 14,973 & 2 & 44,\hspace{1pt}\textbf{47} & 18 & 44,\hspace{1pt}\textit{45} & 18 \\
    \rowcolor{black!10}Plainville & 17,525 & 5 & 22 & 31 & 22 & 31 \\
    Plymouth & 11,671 & 5 & 78 & 31 & 78 & 31 \\
    \rowcolor{black!10}Pomfret & 4,266 & 2 & 50 & \textbf{29} & 50 & \textit{35} \\
    Portland & 9,384 & 1 & 32 & 33 & 32 & 33 \\
    \rowcolor{black!10}Preston & 4,788 & 2 & \textbf{45} & 18 & \textit{42} & 18 \\
    Prospect & 9,401 & 3 & 89 & 16 & 89 & 16 \\
    \rowcolor{black!10}Putnam & 9,224 & 2 & 51 & 29 & 51 & 29 \\
    Redding & 8,765 & 4 & 135 & 26 & \textit{2},\hspace{1pt}135 & 26 \\
    \rowcolor{black!10}Ridgefield & 25,033 & 4 & \textbf{42},\hspace{1pt}111 & \textbf{24},\hspace{1pt}26 & 111,\hspace{1pt}\textit{138} & 26 \\
    Rocky Hill & 20,845 & 1 & 29 & 9 & 29 & 9 \\
    \rowcolor{black!10}Roxbury & 2,260 & 5 & 69 & 32 & 69 & 32 \\
    Salem & 4,213 & 2 & \textbf{34},\hspace{1pt}37 & 20 & 37 & 20 \\
    \rowcolor{black!10}Salisbury & 4,194 & 5 & 64 & 30 & 64 & 30 \\
    Scotland & 1,576 & 2 & 47 & 29 & 47 & 29 \\
    \rowcolor{black!10}Seymour & 16,748 & 3 & 105 & 21,\hspace{1pt}32 & 105 & 21,\hspace{1pt}32 \\
    Sharon & 2,680 & 5 & 64 & 30 & 64 & 30 \\
    \rowcolor{black!10}Shelton & 40,869 & 3,\hspace{1pt}4 & 113,\hspace{1pt}122 & 21 & 113,\hspace{1pt}122 & 21 \\
    Sherman & 3,527 & 5 & 108 & \textbf{30} & 108 & \textit{24} \\
    \rowcolor{black!10}Simsbury & 24,517 & 5 & 16 & 8 & 16 & 8 \\
    Somers & 10,255 & 2 & 52 & 7 & 52 & 7 \\
    \rowcolor{black!10}South Windsor & 26,918 & 1 & \textbf{5},\hspace{1pt}14 & 3 & \textit{11},\hspace{1pt}14 & 3 \\
    Southbury & 19,879 & 5 & 69,\hspace{1pt}131 & 32 & 69,\hspace{1pt}131 & 32 \\
    \rowcolor{black!10}Southington & 43,501 & 1 & \textbf{22},\hspace{1pt}30,\hspace{1pt}80,\hspace{1pt}81 & 16 & 30,\hspace{1pt}80,\hspace{1pt}81,\hspace{1pt}\textit{103} & 16 \\
    Sprague & 2,967 & 2 & 47 & 19 & 47 & 19 \\
    \rowcolor{black!10}Stafford & 11,472 & 2 & 52 & 35 & 52 & 35 \\
\end{tabular}
\newpage
\begin{tabular}{c|c|c|c|c||c|c}
    Town & Pop. & C 2022 & H 2022 & S 2022 & H 2012 & S 2012 \\ \hline
    Sterling & 3,578 & 2 & \textbf{44} & 18 & \textit{45} & 18 \\
    \rowcolor{black!10} Stonington & 18,335 & 2 & \textbf{41},\hspace{1pt}43 & 18 & 43 & 18 \\
    Suffield & 15,752 & 2 & 61 & 7 & 61 & 7 \\
    \rowcolor{black!10}Thomaston & 7,442 & 5 & 76 & 31 & 76 & 31 \\
    Thompson & 9,189 & 2 & 51 & 29,\hspace{1pt}\textbf{35} & 51 & 29 \\
    \rowcolor{black!10}Tolland & 14,563 & 2 & 8,\hspace{1pt}53 & 35 & 8,\hspace{1pt}53 & 35 \\
    Torrington & 35,515 & 1,\hspace{1pt}5 & 63,\hspace{1pt}65 & 8,\hspace{1pt}30 & 63,\hspace{1pt}\textit{64},\hspace{1pt}65 & 8,\hspace{1pt}30 \\
    \rowcolor{black!10}Trumbull & 36,827 & 4 & \textbf{112},\hspace{1pt}122,\hspace{1pt}123,\hspace{1pt}134 & 22 & 122,\hspace{1pt}123,\hspace{1pt}134 & 22 \\
    Union & 785 & 2 & \textbf{52} & 35 & \textit{50} & 35 \\
    \rowcolor{black!10}Vernon & 30,215 & 2 & \textbf{53},\hspace{1pt}56,\hspace{1pt}\textbf{57} & 35 & \textit{8},\hspace{1pt}56 & 35 \\
    Voluntown & 2,570 & 2 & 45 & 18 & 45 & 18 \\
    \rowcolor{black!10}Wallingford & 44,396 & 3 & 85,\hspace{1pt}90,\hspace{1pt}103 & 34 & 85,\hspace{1pt}\textit{86},\hspace{1pt}90,\hspace{1pt}103 & 34 \\
    Warren & 1,351 & 5 & 66 & 30 & 66 & 30 \\
    \rowcolor{black!10}Washington & 3,646 & 5 & \textbf{64} & \textbf{30},\hspace{1pt}32 & \textit{69} & 32 \\
    Waterford & 19,571 & 2 & 38 & 20 & 38 & 20 \\
    \rowcolor{black!10}Watertown & 22,105 & 5 & 68 & 32 & 68 & 32 \\
    Westbrook & 6,769 & 2 & 23,\hspace{1pt}35 & 33 & 23,\hspace{1pt}35 & 33 \\
    \rowcolor{black!10}Weston & 10,354 & 4 & 135 & 26 & 135 & 26,\hspace{1pt}\textit{28} \\
    Westport & 27,141 & 4 & 136,\hspace{1pt}143 & 26 & 136,\hspace{1pt}143 & 26,\hspace{1pt}\textit{28} \\
    \rowcolor{black!10}Wethersfield & 27,298 & 1 & 28,\hspace{1pt}29 & 1,\hspace{1pt}9 & 28,\hspace{1pt}29 & 1,\hspace{1pt}9 \\
    Willington & 5,566 & 2 & 53 & 35 & 53 & 35 \\
    \rowcolor{black!10}Wilton & 18,503 & 4 & \textbf{42} & 26 & \textit{125},\hspace{1pt}\textit{143} & 26 \\
    Winchester & 10,224 & 1 & 63 & 30 & 63 & 30 \\
    \rowcolor{black!10}Windham & 24,425 & 2 & 49 & 29 & \textit{48},\hspace{1pt}49 & 29 \\
    Windsor & 29,492 & 1 & 5,\hspace{1pt}60 & 2,\hspace{1pt}7 & 5,\hspace{1pt}\textit{15},\hspace{1pt}60,\hspace{1pt}\textit{61} & 2,\hspace{1pt}7 \\
    \rowcolor{black!10}Windsor Locks & 12,613 & 1 & 60,\hspace{1pt}\textbf{61} & 7 & 60 & 7 \\
    Wolcott & 16,142 & 5 & 80 & 16 & 80 & 16 \\
    \rowcolor{black!10}Woodbridge & 9,087 & 3 & 114 & 14,\hspace{1pt}17 & 114 & 14,\hspace{1pt}17 \\
    Woodbury & 9,723 & 5 & 66 & 32 & 66,\hspace{1pt}\textit{68} & 32 \\
    \rowcolor{black!10}Woodstock & 8,221 & 2 & 50,\hspace{1pt}\textbf{52} & 35 & 50 & 35 \\
\end{tabular}
\end{document}